\documentclass[12pt,a4paper]{article}

\usepackage[english]{babel}

\usepackage{yhmath}
\usepackage{amsfonts}
\usepackage{amsmath}
\usepackage[single]{accents}

\usepackage{array}
\usepackage{verbatim}
\usepackage{graphicx}
\usepackage{graphics}
\usepackage{subfigure}
\usepackage{epsfig}
\usepackage{tabularx}
\usepackage{mathcomp}
\usepackage{url}

\usepackage{booktabs}
\usepackage[T1]{fontenc}
\usepackage[utf8]{inputenc}

\newcommand{\mG}{\mathcal{G}}

\newcommand{\mT}{\mathcal{T}}

\newcommand{\mV}{\mathcal{V}}

\newcommand{\Sij}{\mathcal{S}_{ij}}
\newcommand{\Skk}{\mathcal{S}_{kk}}

\newcommand{\mbbR}{\mathbb{R}}

\newcommand{\bu}{\mathbf{u}}

\newcommand{\bF}{\mathbf{F}}

\newcommand{\bmG}{\boldsymbol{\mG}}

\newcommand{\bS}{\mathbf{S}}

\newcommand{\bU}{\mathbf{U}}

\newcommand{\bq}{\mathbf{q}}

\newcommand{\bbar}{\overline}

\newcommand{\frho}{\bbar{\rho}}

\newcommand{\wt}{\widetilde}

\newcommand{\fe}{\wt{e}}

\newcommand{\fT}{\wt{T}}

\newcommand{\de}{\partial}
\newcommand{\Div}{\nabla\cdot}
\newcommand{\Grad}{\nabla}

\newcommand{\sigmaij}{\sigma_{ij}}

\newcommand{\RE}{Re}
\newcommand{\MA}{M\hspace{-1pt}a}
\newcommand{\PR}{Pr}

\newcommand{\rvect}{\pmb{r}}
\newcommand{\xvect}{\pmb{x}}

\newcommand{\mean}[1]{\left\langle #1 \right\rangle}

\newcommand\restr[2]{{
		\left.\kern-\nulldelimiterspace 
		#1 
		\vphantom{\big|} 
		\right|_{#2} 
	}}

\newcommand{\partfrac}[2]{\frac{\partial #1}{\partial #2}}
\begin{document}

\title{Dynamical $p-$adaptivity  for LES of compressible flows 
in a high order DG framework}

\author{Antonella Abb\`a $^{(1)}$, Luca Bonaventura$^ {(2)}$\\
Alessandro Recanati $^{(1)}$, Matteo Tugnoli$^{(1)}$ }

\maketitle

 \begin{center}
{\small
(1) Dipartimento di Scienze e Tecnologie Aerospaziali, Politecnico di Milano \\
Via La Masa 34, 20156 Milano, Italy\\
{\tt antonella.abba@polimi.it, alessandro.recanati@mail.polimi.it, matteo.tugnoli@polimi.it}\\
{$ \ \ $ }\\
(2) MOX -- Modelling and Scientific Computing, \\
Dipartimento di Matematica, Politecnico di Milano \\
Piazza Leonardo da Vinci, 20133 Milano, Italy\\
{\tt luca.bonaventura@polimi.it}\\
}
\end{center}

\date{}


\vspace*{0.5cm}


\vspace*{0.5cm}

\pagebreak

\abstract{
	We investigate the possibility of reducing the computational burden of LES models by employing locally and dynamically adaptive polynomial degrees in the framework of a high order DG method.
	A   degree adaptation technique especially featured to be effective for LES applications, that was previously developed by
	the authors and tested in the statically adaptive case,  is applied here in a dynamically adaptive fashion. 
	Two significant benchmarks are considered, comparing the results of adaptive and non adaptive simulations.
	The proposed dynamically adaptive approach allows for a significant reduction of the computational cost of representative LES computation, while allowing to maintain the level of accuracy guaranteed by  LES carried out with constant, maximum polynomial  degree values.
	}


\pagebreak

\section{Introduction}
\label{sec:Introduction}

During the last decades, Large Eddy Simulation (LES) has been extensively studied and has allowed to achieve results comparable to those of Direct Numerical Simulations (DNS) with a significant reduction in computational cost, enabling the use
of high accuracy models to simulate turbulent flows beyond the extremely idealized configurations to which DNS was typically limited. 
While  the increase of the available computing power makes even more fluid flow configurations amenable 
to LES approaches, two interlinked difficulties prevent   LES from being a widespread tool for industrial applications.
 The first major issue is that LES simulations, due to their space and time resolution requirements, 
 are still very computationally expensive. 
The second is that selecting the right spatial resolution for an LES of a unsteady turbulent flow in a complex geometry is not a easy task.
On  one hand, the resolution must be sufficient to solve the equations with adequate accuracy, but limited to contain the computational costs. 
On the other hand, the implicit LES (ILES) commonly employed uses the discretization as the filter, so that the mesh resolution sets implicitly the filter width and the threshold between the resolved scales and the modelled ones.
Moreover, in a turbulent flow in  complex geometry, turbulence characteristics change in time and in space throughout  the computational domain  and cannot be as easily estimated as it was the case in the simple geometries employed for many years in LES studies. 
Therefore, an adaptive LES, where the resolution and the filter width are automatically adapted to the flow conditions, would be of great benefit for the solution of both problems, in order to have an LES with the correct filter sizing and an efficient spatial discretization. The problem of estimating the right resolution is further complicated by the fact that,
as pointed out by the authors in \cite{tugnoli:2017}, using
 standard error estimators to  adapt the resolution is not the optimal approach for 
  LES, if refinement to DNS resolutions is to be avoided, so that
specific strategies must be devised  in this context.

The need of an adaptive LES was first stated by \cite{pope:2004}, but  not many attempts have been made in this direction 
since then.  In the numerical solution of PDEs, however, in the last decades considerable efforts have been devoted to the development of refinement indicators.
Most of these are  based on local estimates of the discretization error, see e.g.
 \cite{antepara:2013, burbeau:2005, kuru:2016, remacle:2003, sagaut:2006, tumolo:2013, tumolo:2015, tumolo:2016, wackers:2014}.
In the LES context, however, simply increasing the resolution in order to decrease the error leads to a DNS solution \cite{mitran:2001, sagaut:2006}, which  is in contrast with the goal of adaptive LES, which consists in adjusting the resolution in order to directly resolve only a prescribed amount of the turbulent scales.
An alternative approach, pursued  e.g. in \cite{hoffman:2005,hoffman:2006,hoffman:2013}, employs indicators aimed at obtaining a good representation of some quantity of interest, rather than of the physical scales of the turbulence.
Adjoint--based adaptation methods proposed by other authors \cite{bassi:2019} have also shown good performance, but entail significant computational costs.

In the present work, we will consider a physically based refinement indicator  proposed by the authors in
\cite{tugnoli:2017, tugnoli:2017a}, which is especially suited for LES. This adaptivity approach is implemented 
in a numerical  framework  based on the Discontinuous Galerkin (DG) method,
that was already presented and validated in \cite{abba:2015}, which allows to extend seamlessly LES concept  to unstructured meshes.
DG methods combine the high order of accuracy and low dissipation/dispersion properties with good parallel performances,
since most of the computations are local to the element and only inter-element boundary fluxes 
must be communicated.  
The  adaptation technique we employ is a $p-$adaptive approach, in which the polynomial degree is varied locally, rather than the mesh size, as in more standard $h-$adaptive
methods. 
DG methods provide an interesting environment for $p-$adaptivity, since they do not require to enforce continuity constraints at the interelement boundaries.
Furthermore, $p-$adaptive techniques are appealing since they allow to correct possible  shortcomings of the computational
mesh, as well as to perform dynamically adaptive simulations without extensive remeshing.
Static polynomial adaptivity was applied to LES in 
\cite{delallaveplata:2019, tugnoli:2017, naddei:2019,  tugnoli:2017a} to efficiently simulate statistically steady phenomena, while
an example of dynamical adaptation is described in \cite{bassi:2019} in an ILES context.

In this paper, we will show how the adaptive method of \cite{tugnoli:2017, tugnoli:2017a} can be applied locally and  dynamically in time, in order to simulate efficiently and accurately transient phenomena.  
A number of time dependent numerical simulations of two different configurations is considered, namely a temporally evolving mixing layer and the interaction of a vortex impinging on a square cylinder with the cylinder wake.
The results obtained show that a significant reduction of the computational cost of LES can be achieved, reducing the number of degrees of freedom employed to values of up to $50\% $ of those required by constant maximum polynomial degree simulations, while maintaining essentially the same level of accuracy.

In Section \ref{sec:Num_meth}, the model equations and numerical method are reviewed.
In Section \ref{sec:Dynamic_Adaptivity}, the dynamical adaptivity approach  we propose is outlined.
In Sections \ref{sec:Mixing_layer},  \ref{sec:Vortex_Cylinder}, results of numerical simulations on two
significant benchmarks are presented.
Finally, in Section \ref{sec:Conclu} some conclusions are drawn and some perspectives of future development
of this approach are discussed.

\section{Model equations and numerical method}
\label{sec:Num_meth}
In this Section, the model equations and the numerical method will be briefly illustrated. For a more detailed description, we refer 
to  \cite{abba:2015}, where this methodology was first introduced and validated for LES applications.
The LES filtered Navier--Stokes equations for compressible flows can be written in  dimensionless form as:
\begin{equation}
\label{eq:compact_form}
\de_t \bU + \Div \bF = 0
\end{equation}
where $\bU=[\frho,\frho\wt{\bu}^T,\frho\fe ]^T$ are the prognostic resolved variables and 
$\bF= \bF^{{\rm ca}} - \bF^{{\rm v}} + \bF^{{\rm sgs}}$ are the fluxes, composed by the advective fluxes
\begin{equation}
\bF^{{\rm a}} (\bU) = \left[
\begin{array}{c}
 \frho\wt{\bu} \\
 \frho\wt{\bu}\otimes\wt{\bu} +
 \frac{1}{\gamma\MA^2}\bbar{p}\mathcal{I} \\
 \frho\wt{h}\wt{\bu}
\end{array}
\right],
\end{equation}
the viscous fluxes
\begin{equation}
\bF^{{\rm v}} (\bU, \Grad \bU) = \left[
\begin{array}{c}
 0 \\
 \frac{1}{\RE}\wt{\sigma} \\
 \frac{\gamma\MA^2}{\RE} \wt{\bu}^T \wt{\sigma}
 - \frac{1}{\kappa\RE\PR} \wt{\bq}
\end{array}
\right]
\end{equation}
and finally the subgrid fluxes
\begin{equation}
\bF^{{\rm sgs}} (\bU, \Grad \bU) = \left[
\begin{array}{c}
 0 \\
 \tau \\
 \frac{1}{\kappa} \mathbf{Q}^{{\rm sgs}}
 +\frac{\gamma\MA^2}{2}\left( 
 \mathbf{J}^{\rm sgs} - \tau_{kk} \wt{\bu}
 \right)
\end{array}
\right].
\end{equation}
Here, $\overline \cdotp$ and $ \wt \cdotp$ represent the grid filter and the Favre filter operators, so that
$\bbar{p}$ denotes the filtered pressure,
$\frho\wt{h} = \frho\wt{e} + \bbar{p}$ the resolved enthalpy and $\wt{\sigma}$ and $\wt{\bq}$ the momentum 
and heat diffusive fluxes, respectively.  
Equations \eqref{eq:compact_form} are completed by the dimensionless state equation for a perfect gas
\begin{equation}
\bbar{p} = \frho \wt{T},
\label{eq:state-eq}
\end{equation}
where $\wt{T}$ is the filtered temperature.
To close the system the constitutive equations must also be specified:
\begin{equation}
\wt \sigmaij = \mu \wt \Sij^d, \qquad
\wt q_i = -\mu\de_i \wt T,
\label{eq:constitutive}
\end{equation}
where the rate of resolved strain tensor is defined as
\begin{equation}
\wt \Sij = \de_j \wt u_i + \de_i \wt u_j \qquad
\wt \Sij^d = \wt \Sij - \frac{1}{3}\wt \Skk\delta_{ij}
\end{equation}
and the dynamic viscosity, according to a power law, is
\begin{equation}
\mu(\wt T) = \wt T^\alpha.
\end{equation}
with $\alpha = 0.7$.
In the Local Discontinuous Galerkin approach \cite{cockburn:1998}, the equations are rewritten as a first order  system 
\begin{eqnarray}
\label{eq:ldg_form}
\de_t \bU + \Div \bF^{\rm a}(\bU) & = & \Div \bF^{\rm v}(\bU,\bmG)
 -\Div \bF^{\rm sgs}(\bU,\bmG) + \bS,   \\
 \bmG - \nabla{\boldsymbol \varphi} & = & 0,
\end{eqnarray}
where ${\boldsymbol \varphi} = [\wt{\bu}^T,\fT]^T$.
A tessellation $\mT_h$ composed of non overlapping tetrahedral elements is defined in the domain $\Omega$ 
over which a discontinuous finite element space $\mV_h$ is defined in the following way
\begin{equation}
\label{eqn:mV_def}
\mV_h = \{ v_h \in L^2(\Omega): v_h|_K \in \mathbb{P}^{p_K}(K), \,
\forall K\in\mT_h \},
\end{equation}
where $\mathbb{P}^{p_K}(K)$ denotes the space of polynomial functions of total degree $p_K$. 
The weak discrete form of equation (\ref{eq:ldg_form}) is derived using test functions belonging to the same space as the 
numerical solution. A modal DG approach is here applied, by using a hierarchical orthonormal polynomial basis for each 
element $K$ in the finite dimensional space $\mV_h$ to represent the numerical approximation of the generic variable 
$a$ expressed as 
\begin{equation}
\label{eq:modal_exp1}
a_h|_K=\sum_{l=0}^{n_\phi(K)}a^{(l)}\phi_{l}^{K},
\end{equation}
where $\phi_{l}^{K}$ are the basis functions on element $K,$
 $a^{(l)} $ are the modal coefficients of the basis functions and $n_\phi(K)+1$ is the number of basis functions required to span 
 the polynomial space $\mathbb{P}^{p_K}(K)$ of degree $p_K$, defined in $\mbbR^3$ as:
 \begin{equation}
 n_\phi(K) =\frac{1}{6}(p_K+1)(p_K+2)(p_K+3)-1
 \label{eq:n_dofs_deg}
 \end{equation}
To compute the advective flux $\{\bF\}^{\rm a}$ the Rusanov flux \cite{leveque:2002} is employed while a centred flux is used for the diffusive fluxes $\{\bF\}^{\rm v}$, $\{\bF\}^{\rm sgs}$ and for the gradient variables flux $\{{\boldsymbol\varphi}\}$.

In the present framework, as explained in detail in \cite{abba:2015}, the filtering operators are built in   the 
DG spatial discretization, so that the LES filtering $\bbar{\cdot}$ is equivalent to the projection onto the employed 
finite dimensional solution subspace 
\begin{equation}
\bbar{a} = \Pi_{\mV_h}a,
\label{eq:filter-bar}
\end{equation}
while Favre filter operator is defined as
\begin{equation}\label{eqn:favre_decomp}
 \bbar{\rho a} = \frho \wt{a}.
\end{equation}
In the simulations presented in this paper,  two different subgrid models have been employed, the classical Smagorinsky 
model, in which the filter width is determined as
\begin{equation}
\Delta(K) = \sqrt[3]{\frac{Vol(K)}{n_\phi(K)+1}}
\label{eq:filtsize}
\end{equation}
and the anisotropic dynamic model \cite{abba:2015, abba:2003}, in which the filter width is included
in the coefficients to be dynamically determined \cite{abba:2017} using the Germano identity \cite{germano:1991}.

\section{A dynamically $p-$adaptive approach}
\label{sec:Dynamic_Adaptivity}

A physically based refinement indicator especially suited for LES has been proposed by the authors in
\cite{tugnoli:2017, tugnoli:2017a}.
This indicator is based on the classical structure function
\begin{equation}
\label{eq:strucfun}
D_{ij} = \mean{\left[u_i(\xvect+\rvect,t)-
	u_i(\xvect,t)\right]\left[u_j(\xvect+\rvect,t)-u_j(\xvect,t)\right]},
\end{equation}
where $\mean{\cdot}$ represents the expected value operator.
The structure function has been widely used to study turbulence statistics and it is known to be directly
related to subgrid stresses \cite{cerutti:1998, cimarelli:2019, germano:2007}. 
Larger values of the structure function calculated inside the element denote a poorly correlated velocity field, 
and the need of a higher 
resolution, while  lower   values denote a highly correlated velocity field, which is an indication of a very well 
resolved turbulent region or laminar conditions, hence the possibility of employing a lower resolution. 
The degree adaptation  indicator is then defined as: 
\begin{equation}
\label{eq:ind_SF}
Ind_{SF}(K) = \sqrt{Q(K)} = \sqrt{\sum_{ij}\left[D_{ij}(K) - D_{ij}
	(K)^{iso}\right]^2}
\end{equation}
where $D_{ij}^{iso}$ is the structure function in isotropic conditions \cite{pope:2000}.

In  \cite{tugnoli:2017}, this procedure was shown to be effective in 
a statically $p-$adaptive framework,   producing accurate results with a significant 
reduction in the computational cost. 
However, static adaptivity presents some limitations, since the resolution is fixed at the beginning of the 
simulation and  constant in time. 
In the simulation of a transient phenomenon, for which a  time resolved solution is sought rather than a statistical average, 
employing a dynamic adaptivity framework
is necessary.  In our implementation of the dynamic degree adaptation,
  indicator \eqref{eq:ind_SF} is computed at runtime at time intervals 
 $\Delta t_i$ and averaged over time. Once the indicator has been averaged on a time interval
 $\Delta t_a $ which is sufficiently larger than  $ \Delta t_i, $ but still small with respect to the time scale of the
 motion of the main turbulent structures,
  a new polynomial distribution is computed based on the indicator values and the flow field approximation is updated,
  by either reducing or increasing the number of degrees of freedom employed in each element.  
    
In this work, the admissible polynomial  degrees ranged from 2 to 4. Two indicator thresholds 
$\epsilon_1 < \epsilon_2$ were employed. Cells with a value of the indicator  \eqref{eq:ind_SF}
 lower than the smaller threshold were assigned
the lowest allowed polynomial degree 2, those with value higher than the largest threshold were
assigned the highest polynomial degree 4 and the intermediate ones were assigned  degree 3. 
Polynomial degrees were only allowed to be increased or decreased by one degree at each adaptation time.
Since the solution is represented by a hierarchical basis, when lowering the polynomial degree,
the contribution associated to the removed modes is simply discarded, while when raising the polynomial degree 
the coefficients  of the newly added modes are initialized to zero, to  be updated by the subsequent time evolution.

 It should be remarked that the   procedure outlined above can potentially lead to unbalances between the computational
 load of  different processors in a parallel run.  The implementation of a full dynamic load balancing is beyond the scope 
of the present work, whose focus is mostly on the assessment of the  accuracy of the above  methodology. As a consequence,
in all the numerical experiments presented below, the computational load among the processors 
is approximately balanced at the beginning of the simulation only, which is clearly suboptimal with respect to full
parallel efficiency. Indeed,
the results will not in general be assessed based of the CPU time required, but rather on the total number of the degrees
of freedom employed in the adaptive versus non adaptive simulations. However, 
it must be noted that,t even with this suboptimal configuration, the adaptive simulation always led to a net reduction in computational effort. In order to assess the effective reduction in the 
CPU time required, further   work on code optimization is needed, along with the   development and application of load 
balancing approaches such as those employed e.g. in \cite{bassi:2019, wang:2019}.

\section{Temporally evolving mixing layer}
\label{sec:Mixing_layer}

In a first assessment of the adaptive LES approach outlined in the previous sections,
an isothermal time developing mixing layer was simulated.
The temporal mixing layer represents an interesting test case for LES,   due to its simplicity and advantages from the computational
viewpoint, as well as  the complexity of the physics related to the mixing, see e.g. the discussions in \cite{foysi:2010, sandham:1991, vreman:1995a, yang:2004}.
The commonly employed configuration to represent a mixing layer flow was employed, imposing  
periodic boundary conditions on the mean flow and in the span wise directions.
Thanks to the periodicity in the stream wise direction, uncertainties related to the imposition of inlet and outlet boundary conditions are avoided. 
In the simulations  presented here, the anisotropic dynamic subgrid model \cite{abba:2015, abba:2003} has been used, in which the filter width is included in the coefficients to be dynamically determined. A further application of this approach is presented in \cite{abba:2017}.

\begin{figure}
	\centering
	\includegraphics[width=0.5\textwidth]{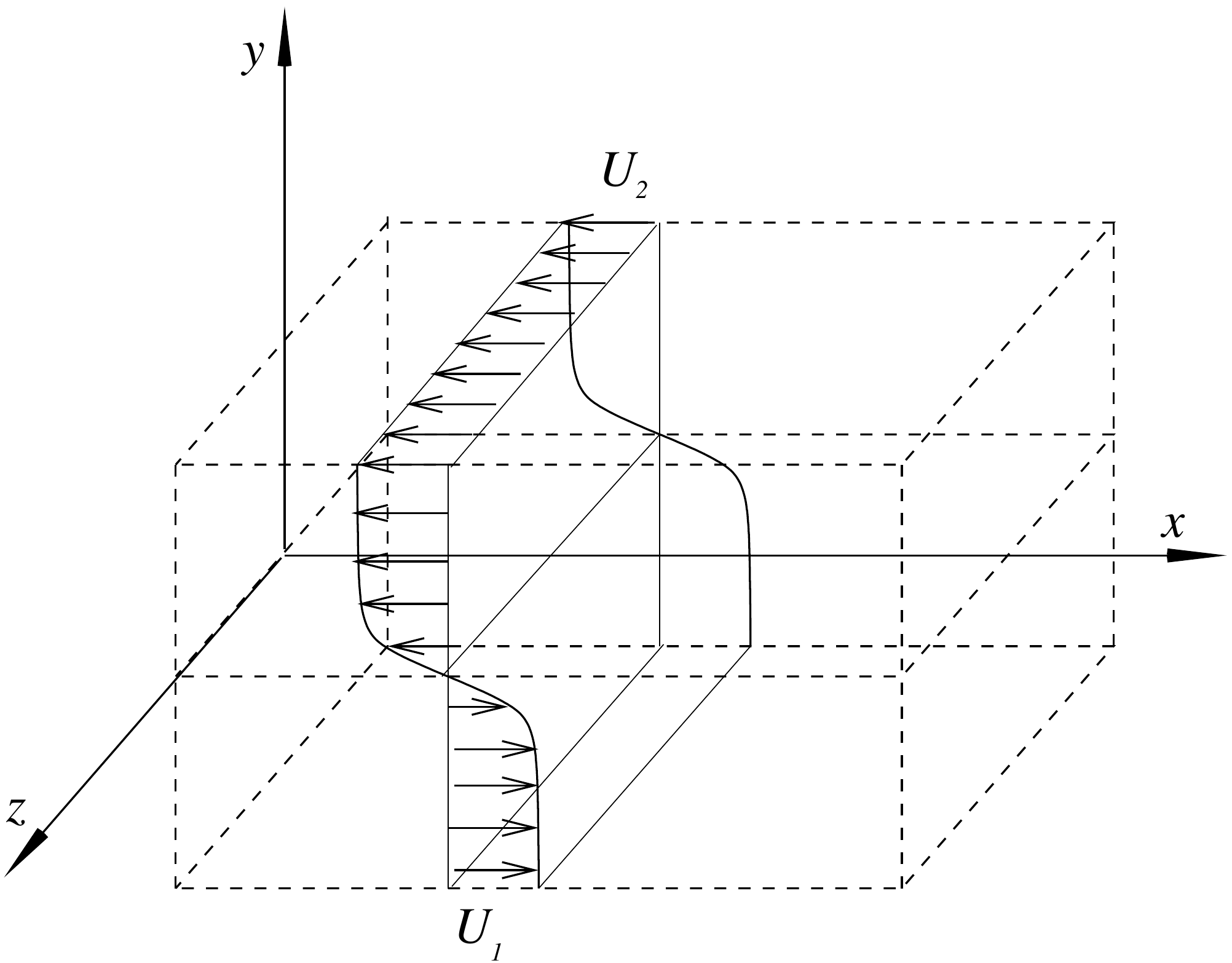}
	\caption{Sketch of the mixing layer configuration.}
	\label{fig:ml_geometry}
\end{figure}

A sketch of the flow configuration is shown in Figure \ref{fig:ml_geometry}. 
The mixing layer is characterized by two parallel flows with different velocities $U_1$ and $U_2. $ The convection velocity of the isothermal mixing layer, defined as $U_c = (U_1+U_2)/2$, is  the velocity that transports the cortical structures at the centreline. In the configuration we considered, the convection velocity is assumed to be zero, so that the eddies do not travel inside the domain by means of the flow, but only move by mutual interaction. 
As a consequence, $U_1 = -U_2$ and   $|U_1|=|U_2|=(U_1-U_2)/2$.
The reference frame is chosen so that the $x$ axis is aligned with the main direction of the flow centred at the middle of the domain, while the $y$ and $z$ axis denote transverse and lateral directions, respectively.
The initial vorticity thickness, defined as $$\delta_{\omega r}= \frac{U_1 - U_2}{\max(dU/dy)}$$ is chosen as reference length for the non-dimensionalization. 
The  initial velocity jump $\Delta U_r = (U_1-U_2)$ is used as reference velocity, the initial uniform density $\rho_r$ and temperature $T_r$ are used as reference density and temperature, respectively. 
The fundamental dimensionless groups are then
 $$ Re= \frac{\rho_r \Delta U_r \delta_{\omega r}}{\mu_r}=400,   \quad Pr = \frac{\mu_r c_p}{k}=0.71, \quad  
 Ma= \frac{\Delta U_r}{\sqrt{\gamma R T_r}}=0.2. $$
The temporally evolving mixing layer develops from a specified initial condition. In
these computations, we use a hyperbolic tangent as the base velocity profile for the longitudinal component
\begin{equation}
 U(y) = \frac{U_1 + U_2}{2} + \frac{U_1 - U_2}{2} \tanh (2y) 
\end{equation}
where $U_1=0.5$ and $U_2=-0.5$ so that $U_c=0$. 
Similarly to \cite{golansky:2005,fortune:2004}, a 3D incompressible disturbance is added to the base velocity profile to initiate the transition process. 
It consists of harmonic disturbances expressed as
\begin{eqnarray}
u'& = & A e^{-\sigma y^2}  \frac{\sigma}{\pi} L_x y \left[ \sin\left(\frac{16 \pi}{L_x} x \right) +
\frac{1}{8} \sin\left(\frac{8 \pi}{L_x} x \right) + \frac{1}{16} \sin\left(\frac{4 \pi}{L_x} x \right) \right] \nonumber \\
v'& = & A e^{-\sigma y^2}  \left[ \cos\left(\frac{16 \pi}{L_x} x \right) +
\frac{1}{8} \cos\left(\frac{8 \pi}{L_x} x \right) + \frac{1}{16} \cos\left(\frac{4 \pi}{L_x} x \right) \right] \nonumber \\
 w'& = & A e^{-\sigma y^2}   \frac{\sigma}{\pi} L_x y \left[ \sin\left(\frac{16 \pi}{L_x} x \right) +
\frac{1}{8} \sin\left(\frac{8 \pi}{L_x} x \right) + \frac{1}{16} \sin\left(\frac{4 \pi}{L_x} x \right) \right] \nonumber \\
& & \left[ \cos\left(\frac{16 \pi}{L_x} x \right) +
\frac{1}{8} \cos\left(\frac{8 \pi}{L_x} x \right) + \frac{1}{16} \cos\left(\frac{4 \pi}{L_x} x \right) \right] \nonumber
\end{eqnarray}
where $A = 0.025(U_1-U_2 )$ and the decay in $y$-direction is governed by $\sigma = 0.05$.
Moreover, a 3D white noise perturbation  is added to the initial velocity field,
generated  for each mesh node by a logistic map rescaled to take values on  $[-A,A]$ and multiplied by the attenuation factor
$e^{-\sigma y^2}.$  Uniform pressure $p=1$ and density
$\rho=1$ are used to initialize the simulation.

The longitudinal dimension  of the computational domain must be large enough to allow at least
for the merging of two principal vortical structure, so it must be
taken as a multiple of the wavelength $\lambda_a$ characterizing the most unstable perturbation, which
according to the analysis in \cite{sandham:1991} is given in this context by
 $\lambda_a = 7.66. $   In the present simulation, the size in the streamwise direction has been chosen 
 $L_x = 8 \times \lambda_a = 60.8,$ so that up to three vortices are  allowed to merge.
  In the normal direction, the computational 
 domain extends for $-30 \leq y \leq 30,$ while in spanwise direction $L_z=30.4$.
 Periodic boundary conditions are applied in the statistically homogeneous $x$ and $z$ directions, while a sponge layer 
 is employed at the top and bottom boundaries,  so that the effective size of the computational domain
in the $y$ direction is equal to $80$.

 The computational mesh is built first subdividing the domain in $N_{hx}\times N_{hy} \times N_{hz} = 20 \times 36 \times 10 $ 
 hexahedra. Each hexahedron is then split into $6$   tetrahedra, yielding a total   of $43200$   tetrahedral elements.
The actual resolution depends on the local polynomial order and can be computed as
$$
\Delta_i = \frac{L_i}{N_{hi}\sqrt{6 N_q}^3}.
$$
For $p=4,$ we get
$\Delta_x=5.11, \quad \Delta_y = 0.37, \quad \Delta_z = 0.5$.

A preliminary study of the mixing layer in the present configuration was presented in \cite{recanati:2019}, where the  sensitivity of the results  to the choice of spatial resolution and subgrid scale model was checked. The objective of the present work is to test the 
 $p-$adaptivity technique in a time evolving flow. For this purpose,
 the results obtained with polynomial degree  dynamically adapting in space and time are compared 
with the results obtained with constant fourth and third degree polynomials.

To set the values of the adaptation thresholds $\epsilon_1, \epsilon_2, $ the structure function indicator was evaluated on the  velocity field obtained with constant $p=4 $ at the final dimensionless time $t=150. $ Thresholds were  then  chosen in order to obtain, for that field, an adapted number of degrees of freedom (dofs) lower than  $80\% $ of the corresponding number of dofs in the constant $p=3 $ case.
The resulting threshold values are then $\epsilon_1=10^{-6}$ and $\epsilon_2= 2\times10^{-3}$.
The dynamical $p-$adaptation was carried out by computing the indicator value at intervals
$\Delta t_i=0.0375 $ and averaging them over intervals  
 $\Delta t_a= 0.075, $ while the value  $\Delta t=0.0075 $ was used for time integration.

\begin{figure}
	\centering
	\subfigure[]{\resizebox*{3.cm}{!}
	{\includegraphics{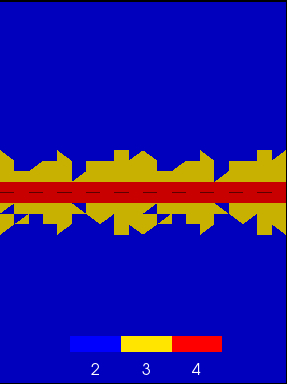}}}
	\hfill
	\subfigure[]{\resizebox*{3.cm}{!}
	{\includegraphics{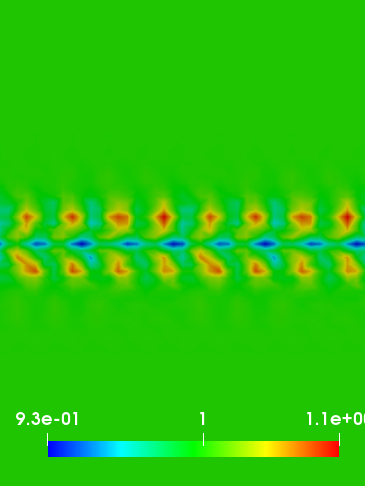}}}
	\hfill
	\subfigure[]{\resizebox*{3.cm}{!}
	    {\includegraphics{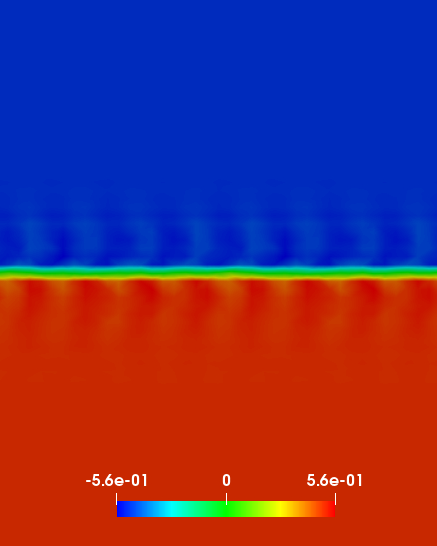}}}\\
	\subfigure[]{\resizebox*{3.cm}{!}
		{\includegraphics{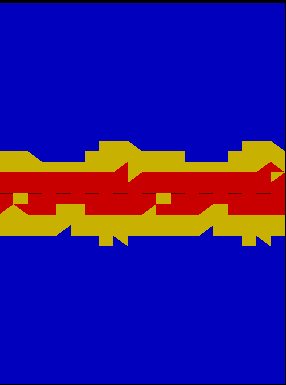}}}
	\hfill
	\subfigure[]{\resizebox*{3.cm}{!}
	{\includegraphics{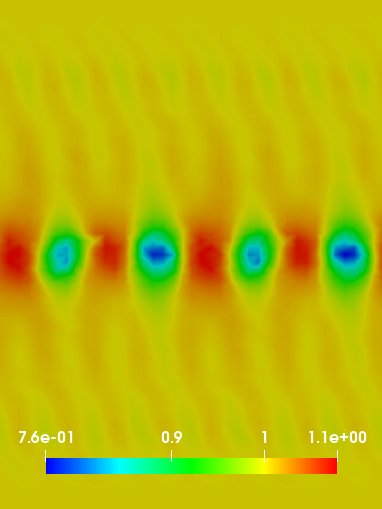}}}
	\hfill
	\subfigure[]{\resizebox*{3.cm}{!}
	{\includegraphics{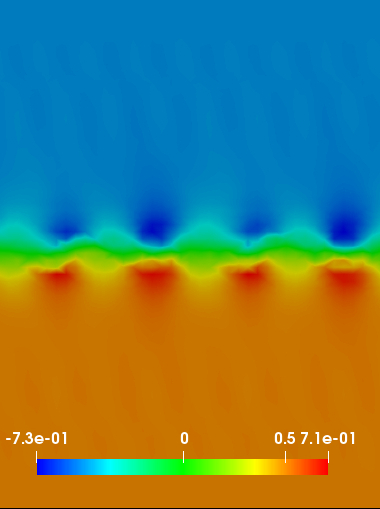}}}\\
	\subfigure[]{\resizebox*{3.cm}{!}
		{\includegraphics{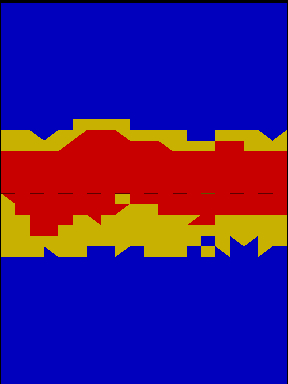}}}
	\hfill
	\subfigure[]{\resizebox*{3.cm}{!}
	    {\includegraphics{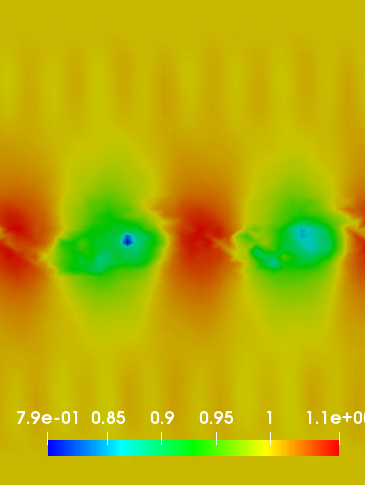}}}
	\hfill
	\subfigure[]{\resizebox*{3.cm}{!}
	    {\includegraphics{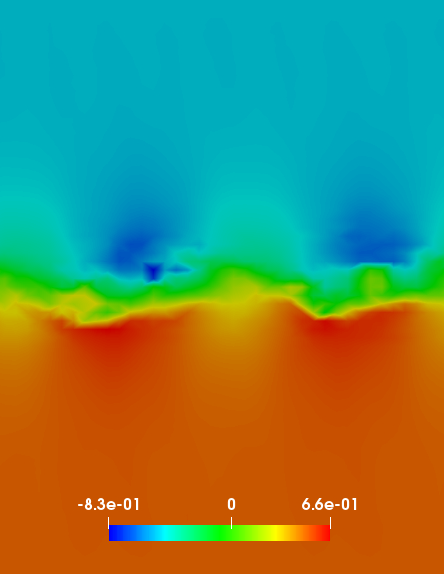}}}\\
	\subfigure[]{\resizebox*{3.cm}{!}
		{\includegraphics{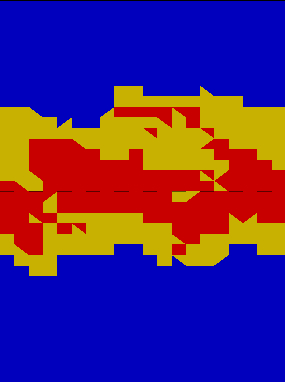}}}
	\hfill
	\subfigure[]{\resizebox*{3.cm}{!}
		{\includegraphics{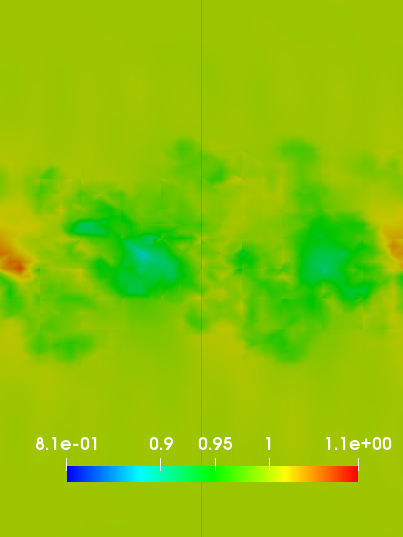}}}
	\hfill
	\subfigure[]{\resizebox*{3.cm}{!}
	    {\includegraphics{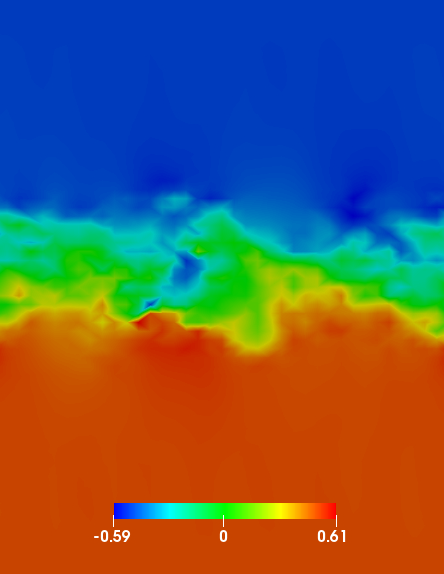}}}
	\caption{Results of the $p-$adaptive simulation of the temporal mixing layer plotted in the plane $z=0$ 
	at times $t=4, 63,126,150$:
	 polynomial degree distribution (left); density (centre); longitudinal velocity component (right).}
	\label{fig:ml_ux_pdeg}
\end{figure}

\begin{figure}
	\centering
	\includegraphics[width=0.5\textwidth]{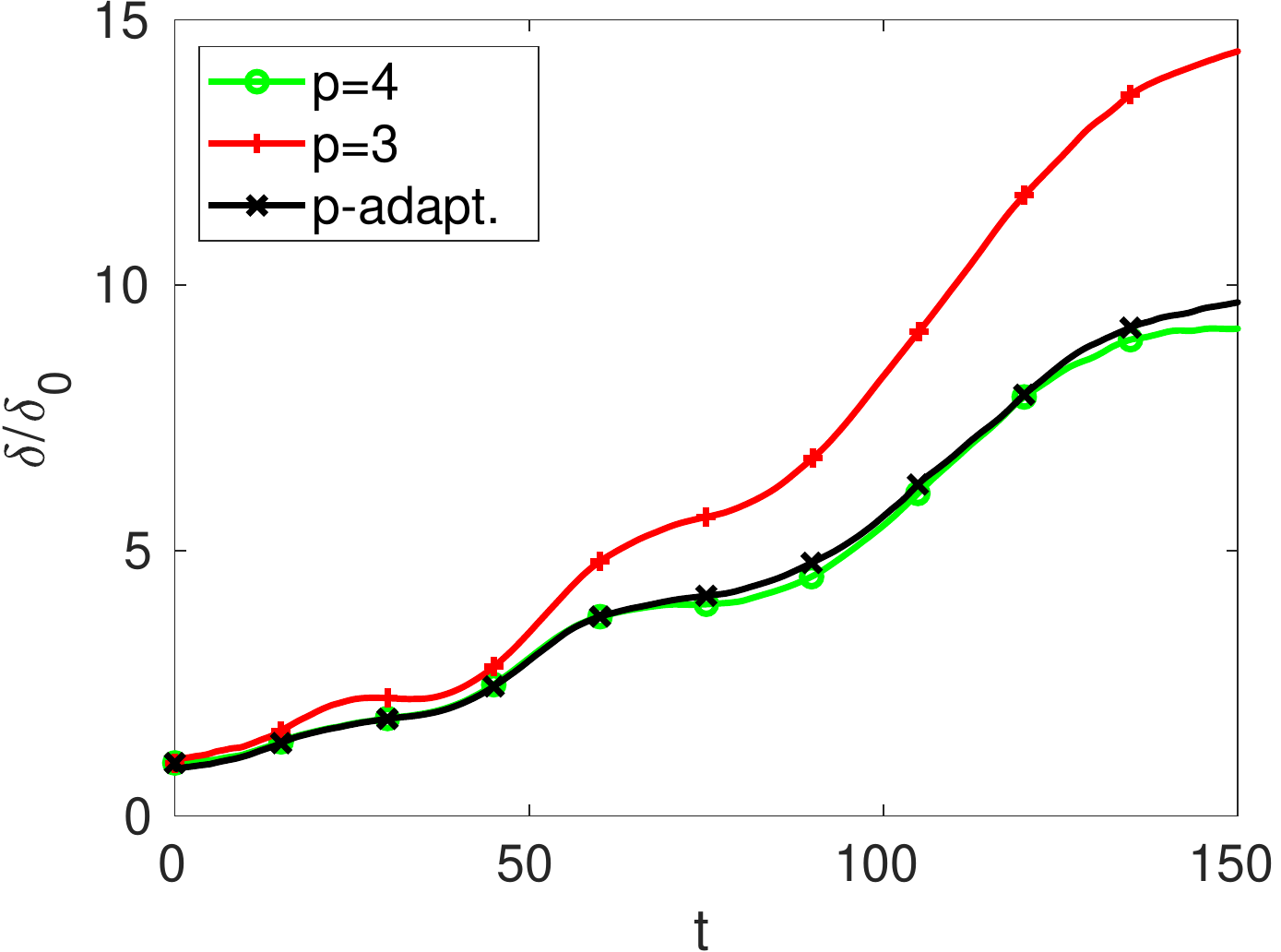}
	\caption{Growth of the momentum thickness, normalised by its initial value, versus time, 
	for the $p-$adaptive case and for uniform degree  cases $p=4$, $p=3$.}
	\label{fig:ml_delta}
\end{figure}

The polynomial degree distribution, the density and the streamwise component of the velocity, at different instants,  in the plane $z=0$, are represented in Figure \ref{fig:ml_ux_pdeg}.
In the initial condition, the elements with highest degree $p=4$ are concentrated along
 the middle of the domain, where high resolution is required.
Then, the region with higher polynomial degree extends in space during time, 
adapting to the merging of the vortices, to the diffusion of turbulent structures and to the growth of the layer thickness.
An important quantity for the characterization of the mixing layer is the momentum thickness, defined as
\begin{equation}
 \delta_\theta(t) = \int \overline\rho (\frac{1}{2}- \tilde u)(\frac{1}{2} + \tilde u) dy.
 \label{eq:ml_delta_theta}
\end{equation}
The time evolution of this quantity is represented in Figure \ref{fig:ml_delta}. 
Notice that slope variations in the time evolution are associated to the merging of the vortices.
A difference in the  behaviour of the $p=3$ simulation with respect to the $p=4$ one is evident after the second merging, while the $p-$adaptive simulations well reproduces the  $p=4$ results.

\begin{figure}  
	\centering
	\subfigure[$t=60$]{\resizebox*{6cm}{!}
	    {\includegraphics{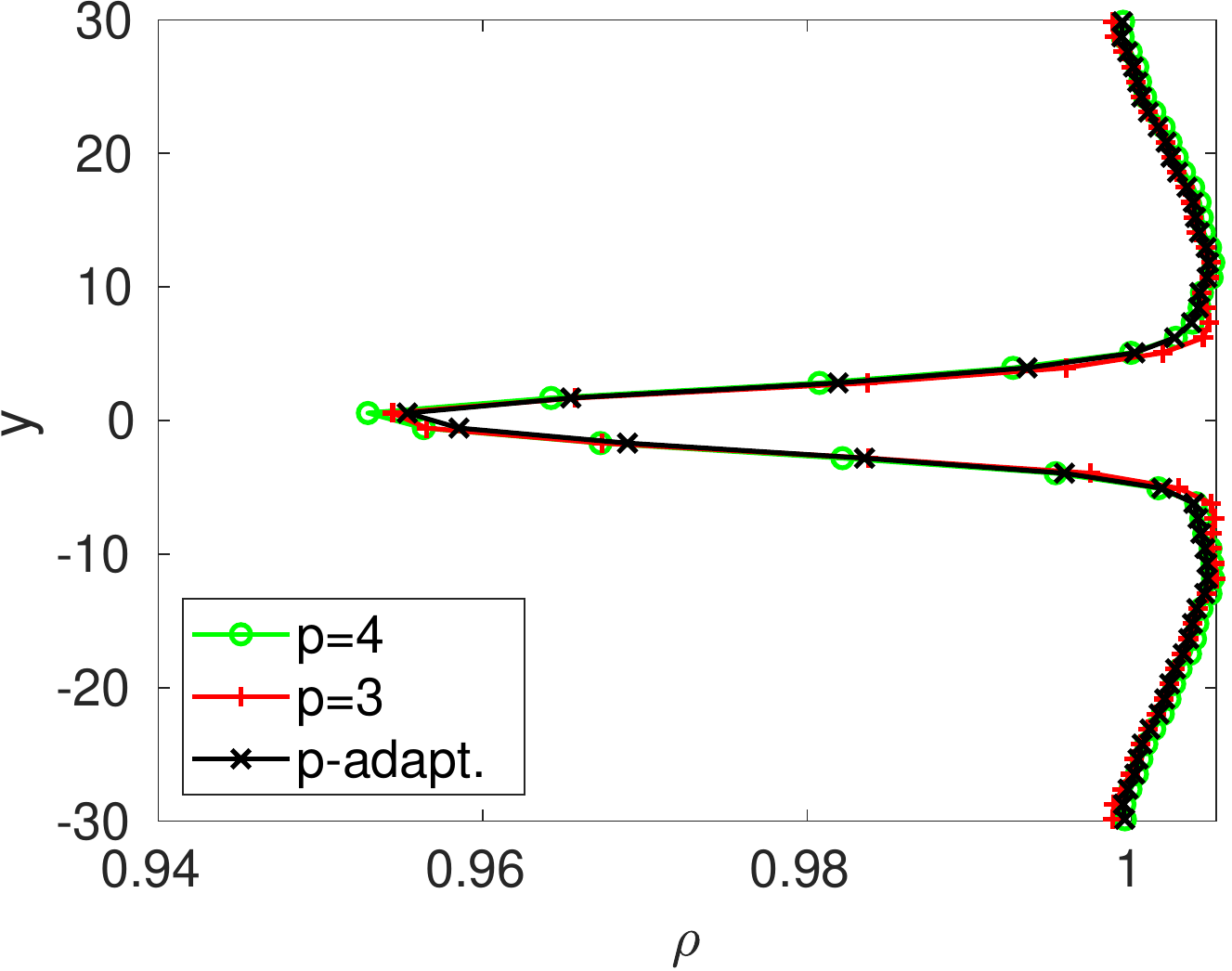}}}
	\subfigure[$t=60$]{\resizebox*{6cm}{!}
		{\includegraphics{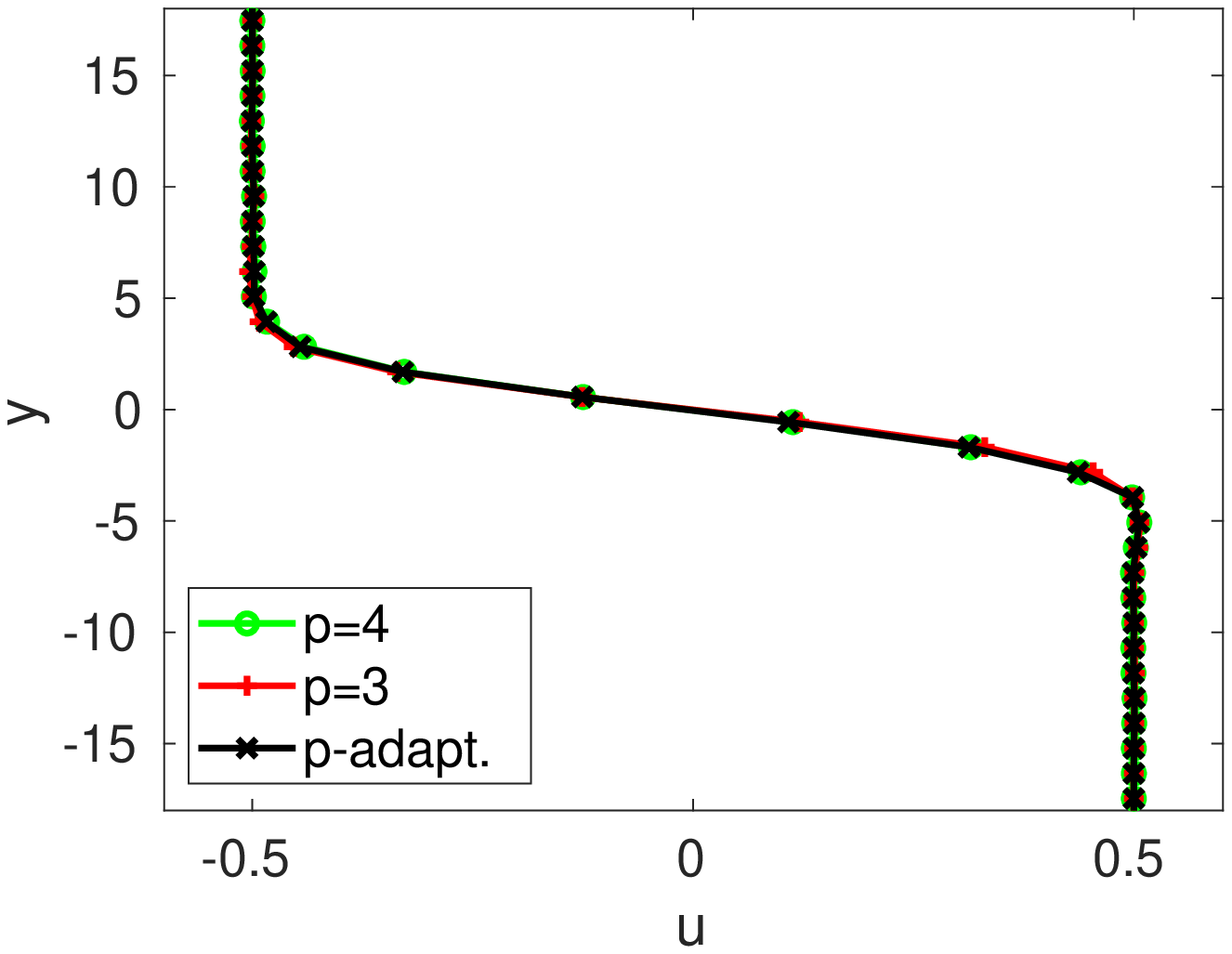}}}\\
	\subfigure[$t=105$]{\resizebox*{6cm}{!}
	    {\includegraphics{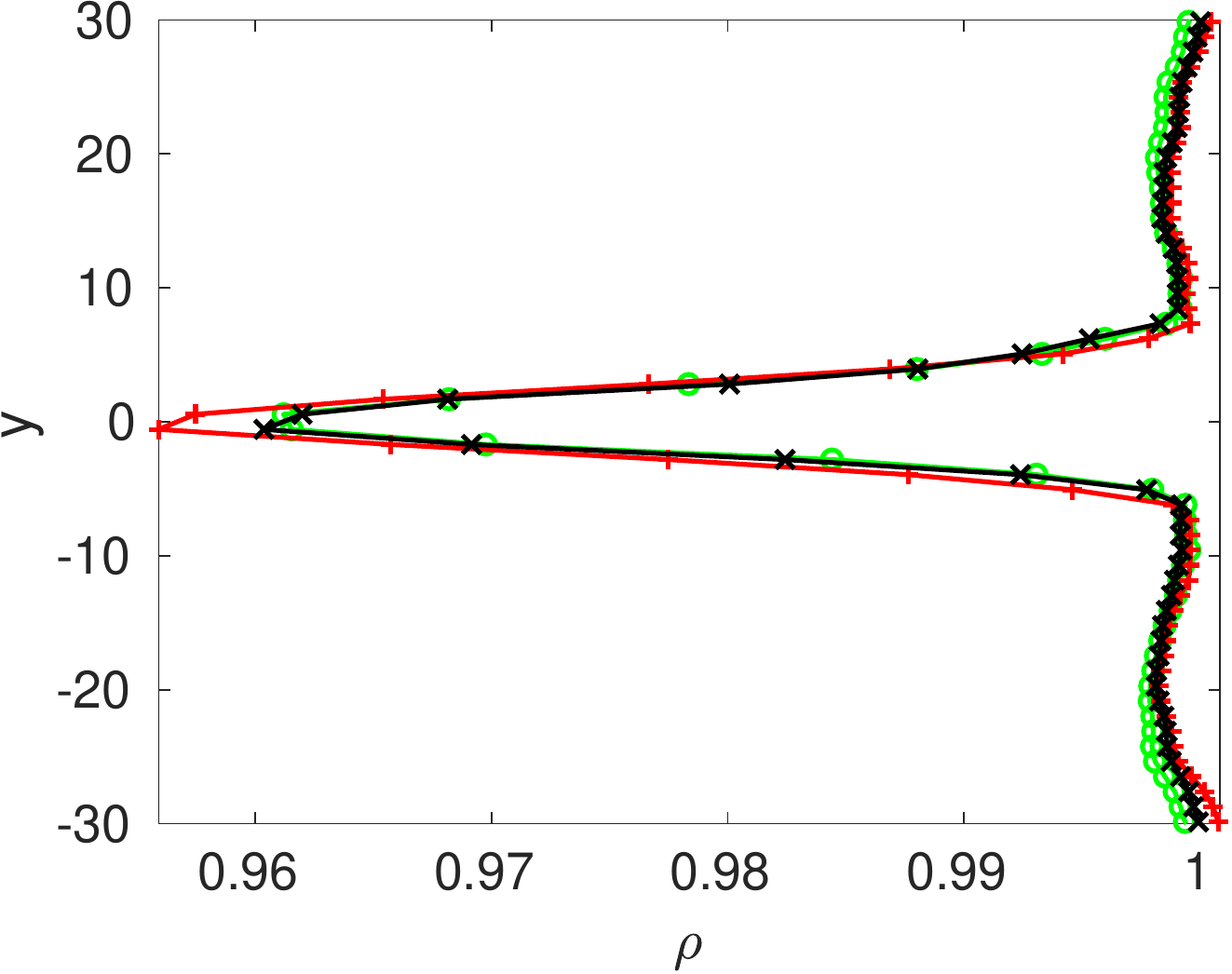}}}
	\subfigure[$t=105$]{\resizebox*{6cm}{!}
		{\includegraphics{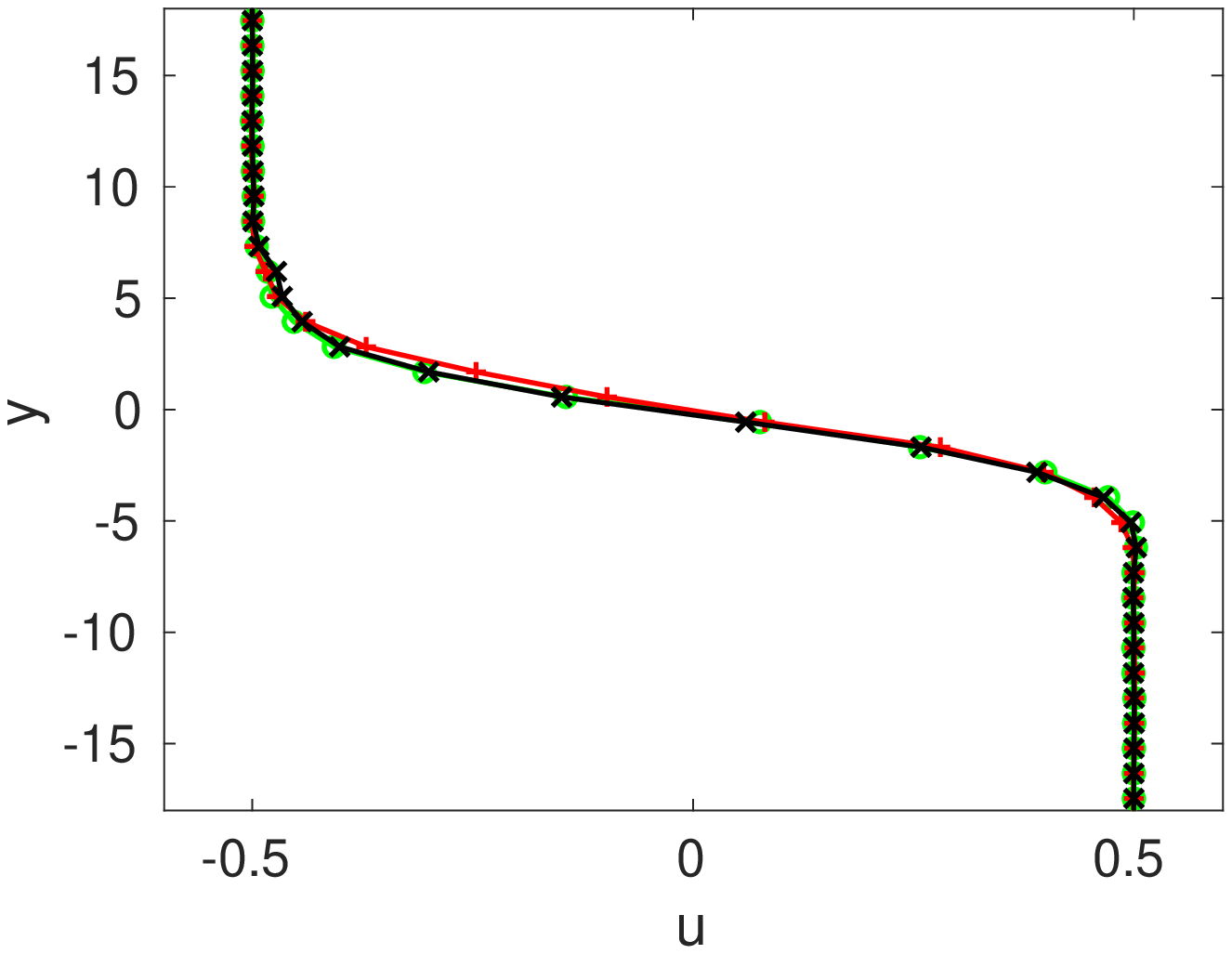}}}\\
	\subfigure[$t=150$]{\resizebox*{6.1cm}{!}
	    {\includegraphics{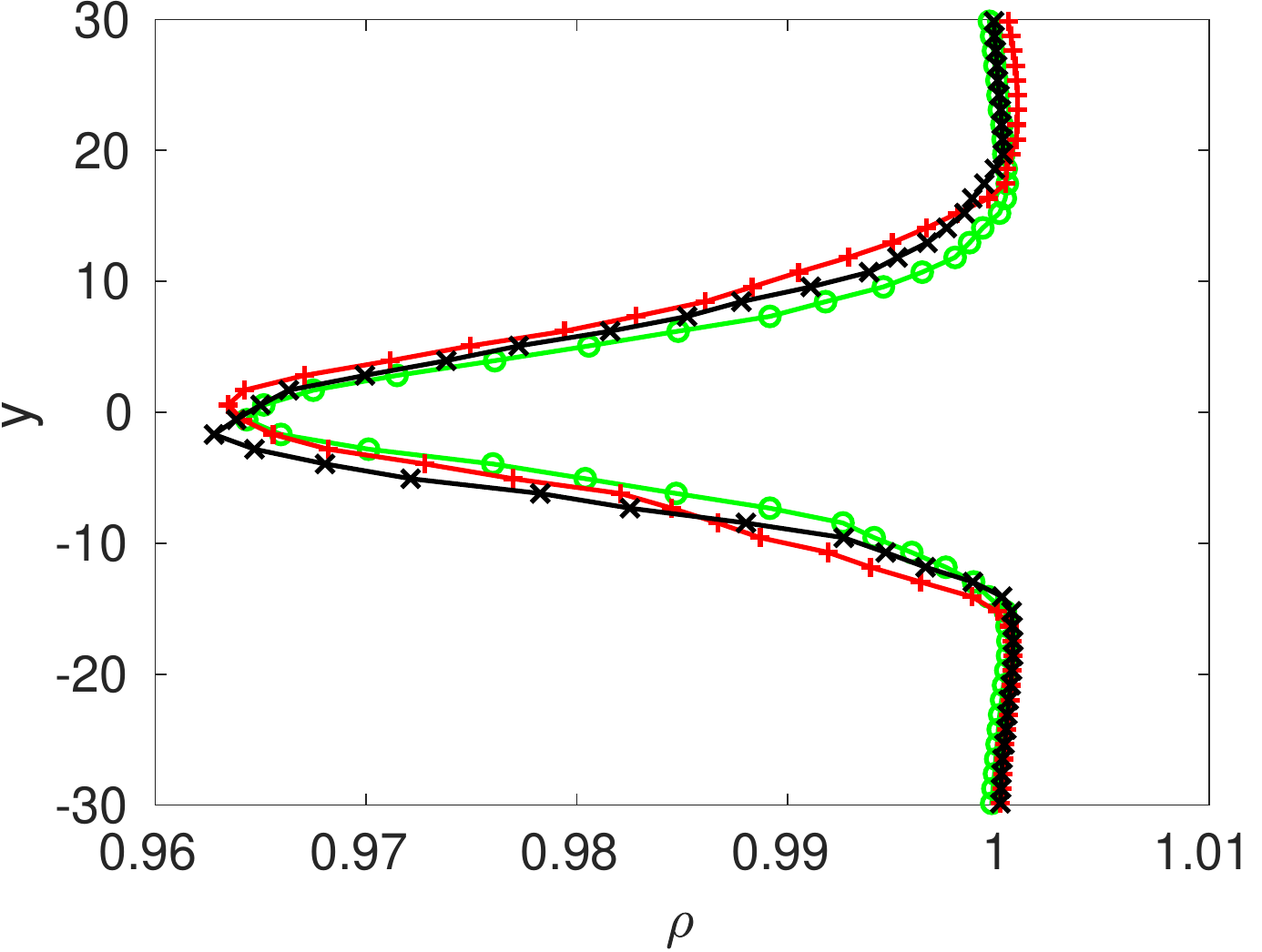}}}
	\subfigure[$t=150$]{\resizebox*{6cm}{!}
		{\includegraphics{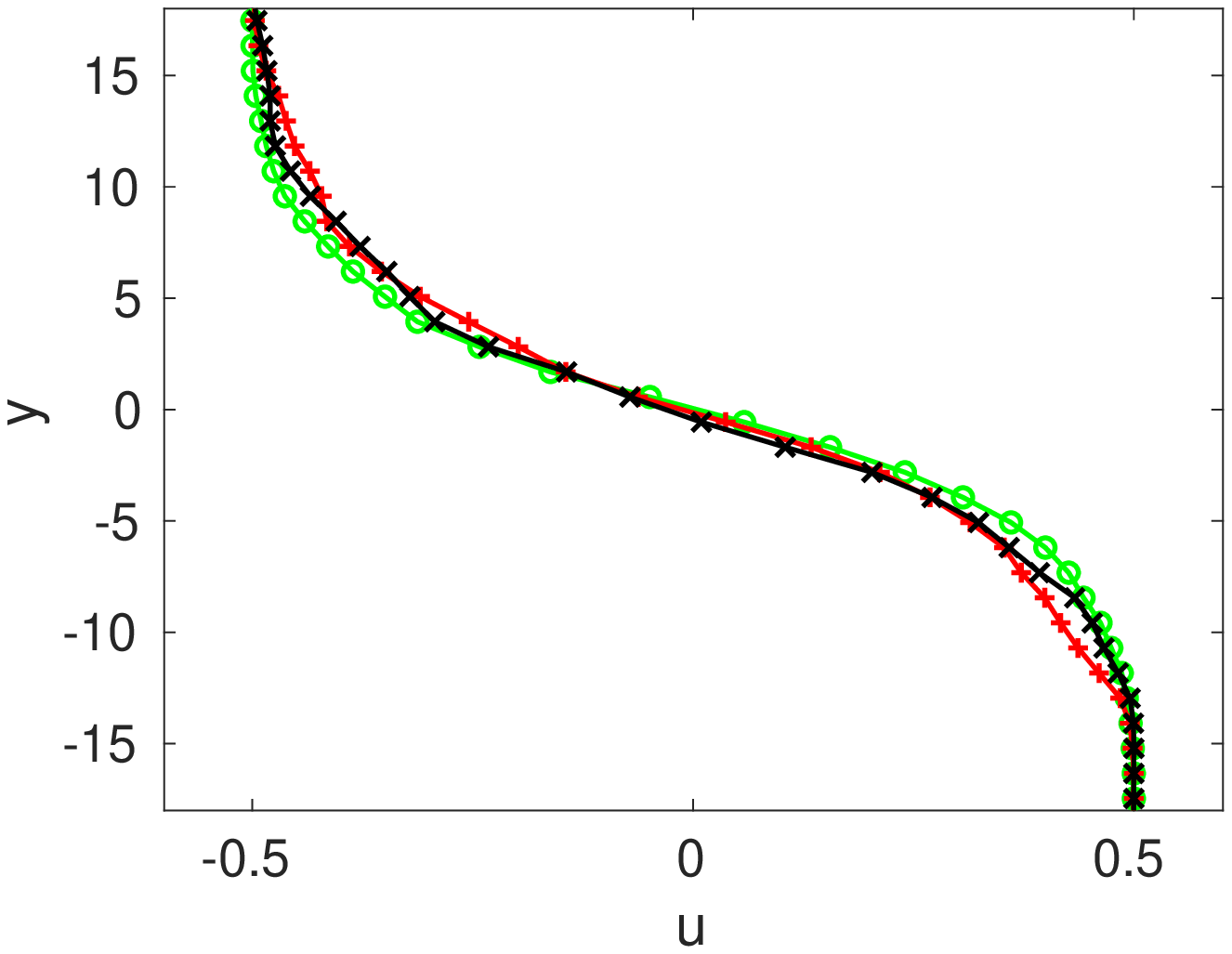}}}
	\caption{Mean profiles for density (left) and longitudinal velocity component (right) for the mixing layer 
	at different times $t$.}
	\label{fig:ml_mean_prof}
\end{figure}

\begin{figure}
	\centering
	\subfigure[$\tau_{11}$]{\resizebox*{6cm}{!}
	    {\includegraphics{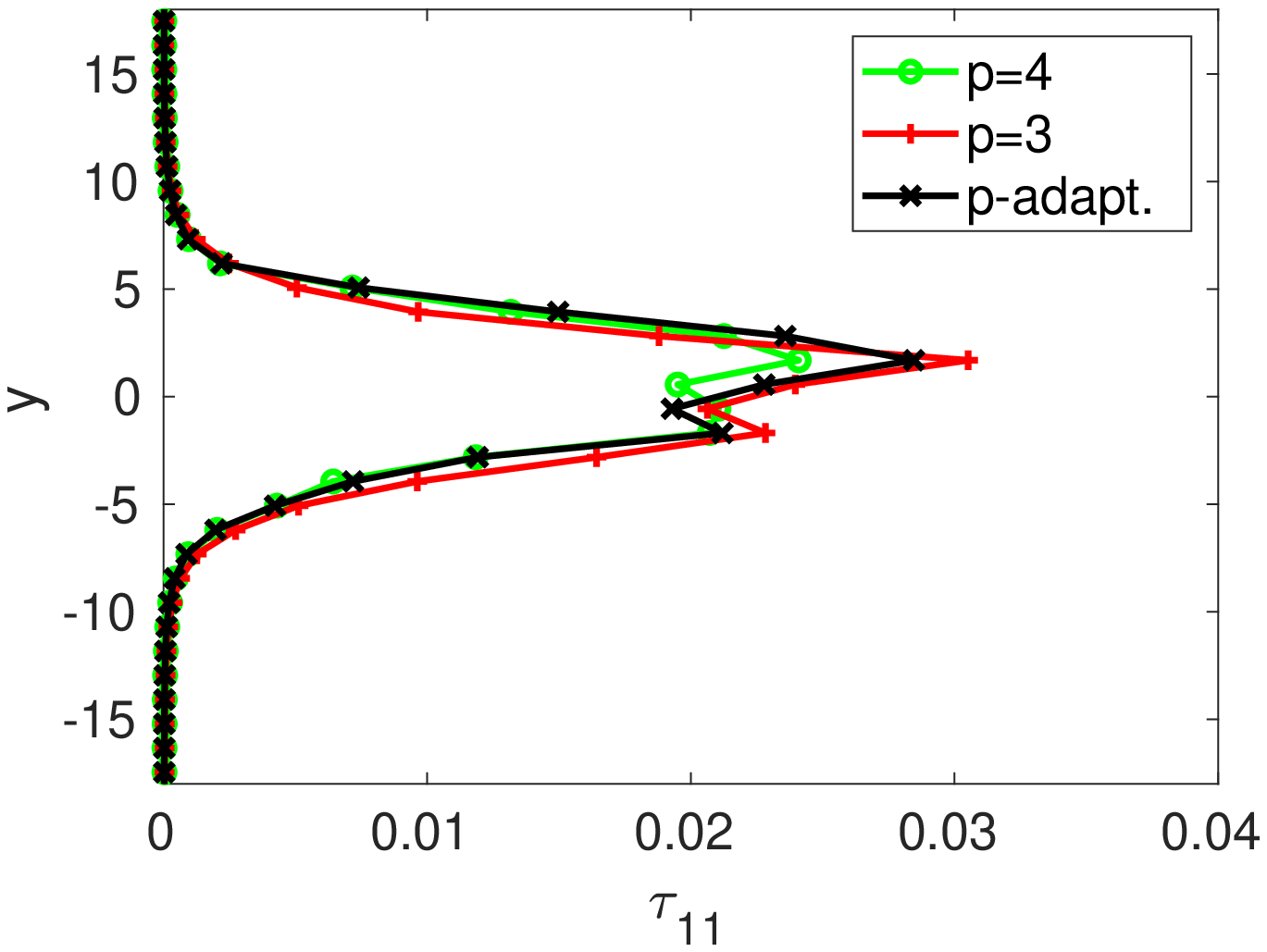}}
	    \label{fig:ml_mean_stress_a}}
	\subfigure[$\tau_{22}$]{\resizebox*{6cm}{!}
	    {\includegraphics{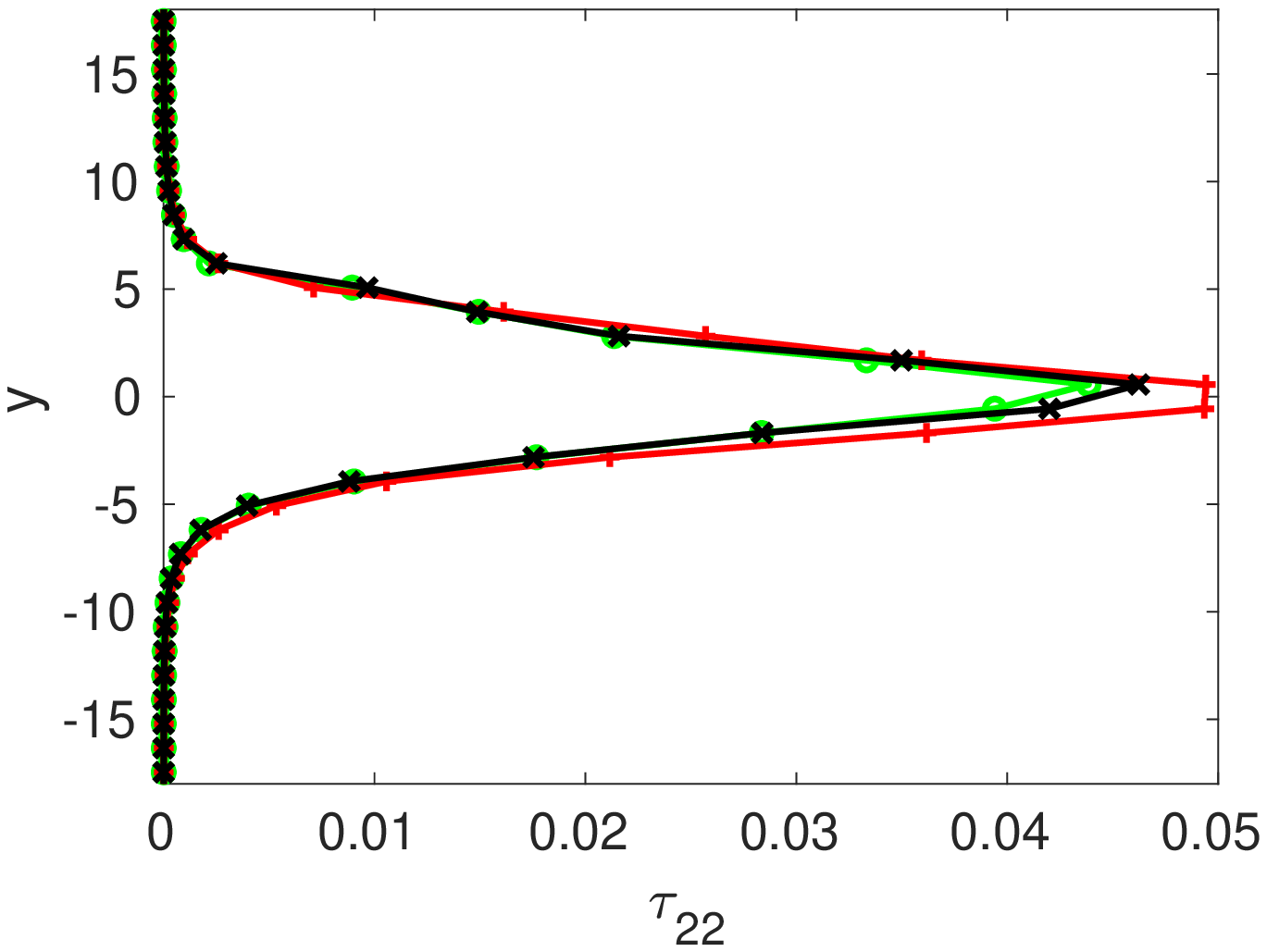}}
	  \label{fig:ml_mean_stress_b} } \\
	\subfigure[$\tau_{33}$]{\resizebox*{6cm}{!}
	    {\includegraphics{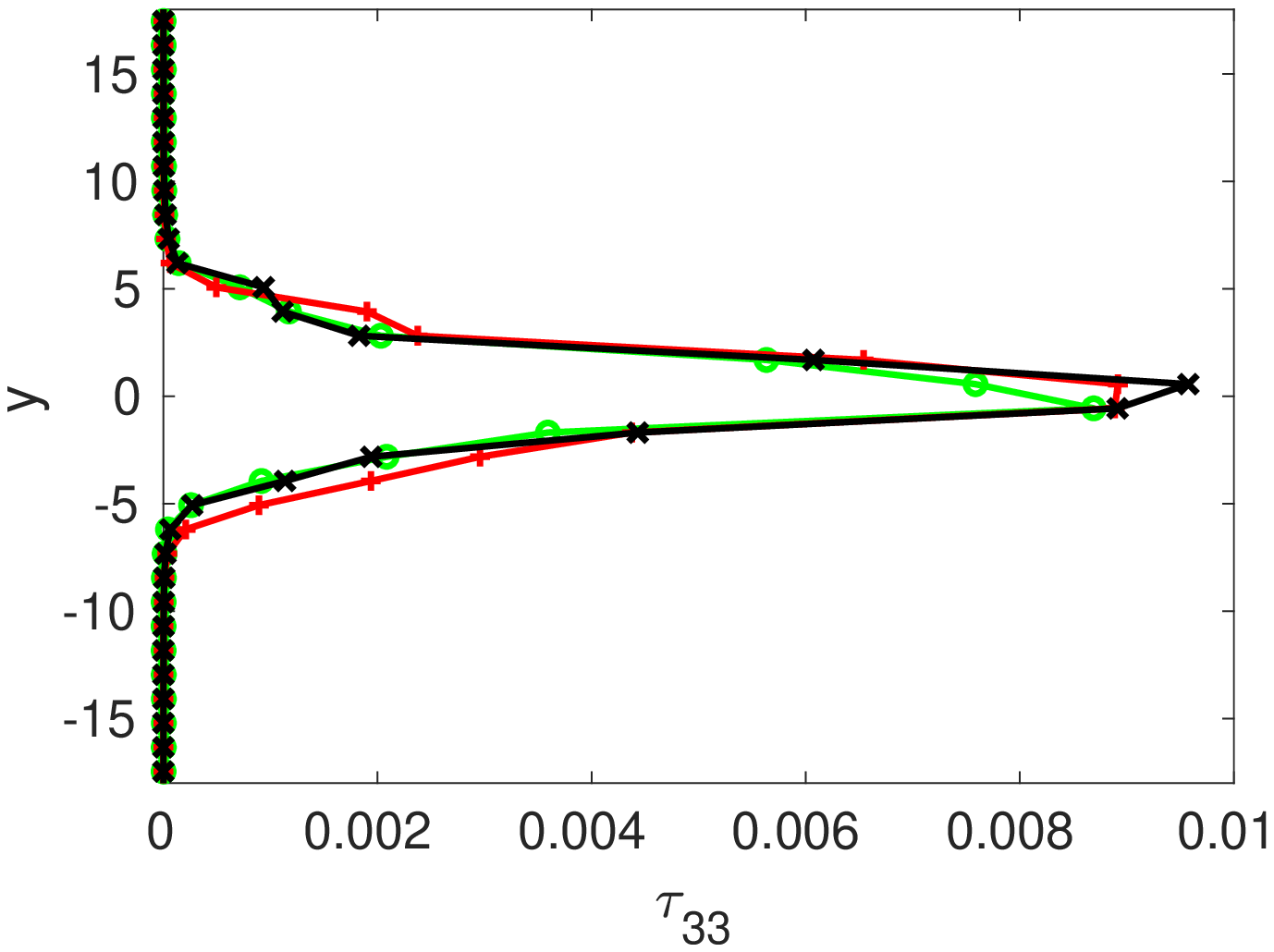}}
	    \label{fig:ml_mean_stress_c}}
	\subfigure[$\tau_{12}$]{\resizebox*{6cm}{!}
		{\includegraphics{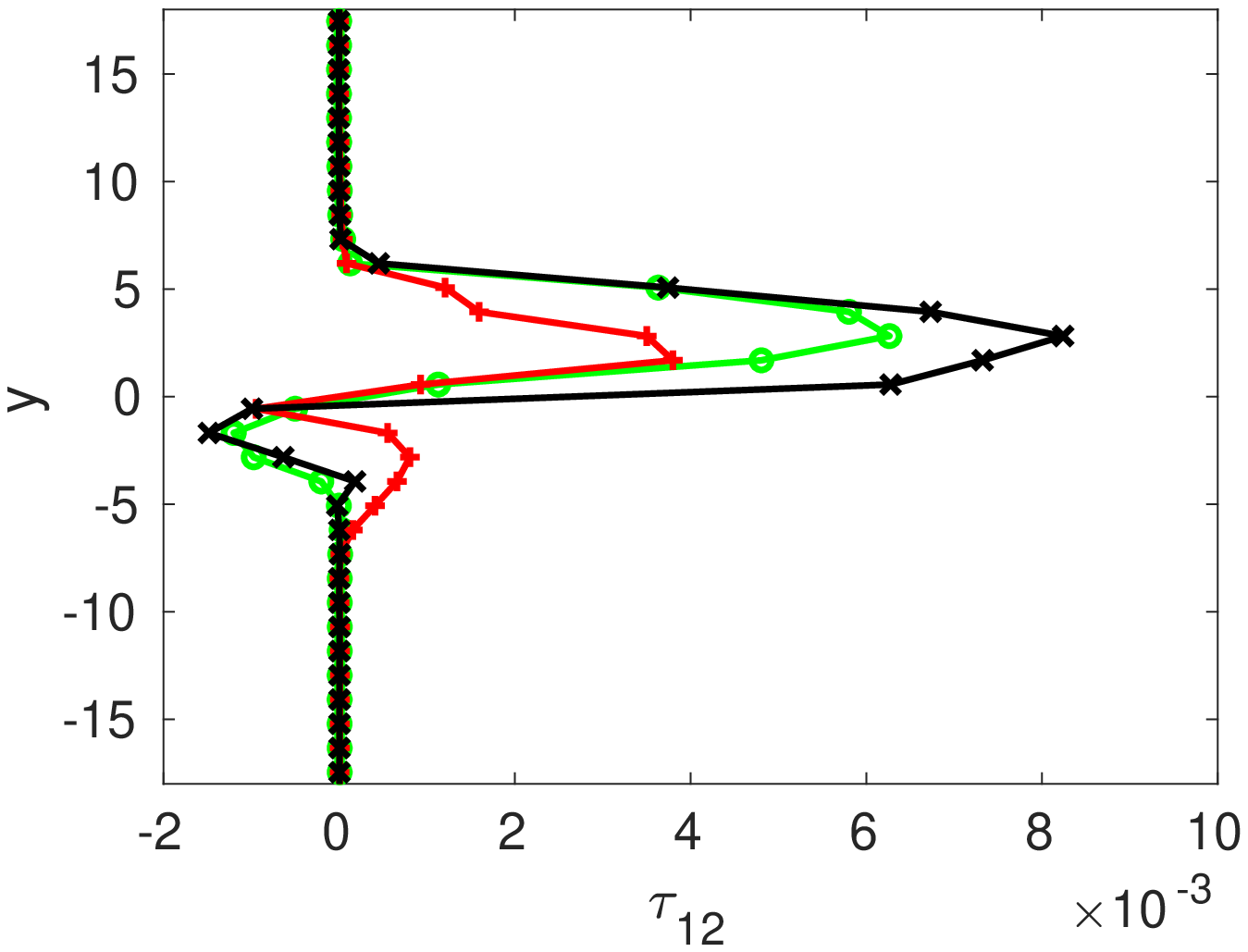}}
		\label{fig:ml_mean_stress_d}}
	\caption{Mean profiles of the total turbulent stresses for the mixing layer at time $t=150$.}
	\label{fig:ml_mean_stress}
\end{figure}

No relevant differences between the results obtained using adaptive or constant polynomial degree distribution are observed in
the mean density and velocity profiles, shown in
Figure \ref{fig:ml_mean_prof}, nor in
 the mean normal stresses profiles, shown in Figures \ref{fig:ml_mean_stress_a}--\ref{fig:ml_mean_stress_c}.
Instead, looking at the shear stresses in Figure \ref{fig:ml_mean_stress_d}, although the peak is slightly overestimated, the $p-$adaptive solution is in good agreement with that of the $p=4 $ case. 
From the same picture, the more dissipative character of the $p=3 $ solution is also apparent.

\begin{figure}
	\centering
	\subfigure[]{\resizebox*{6cm}{!}
	    {\includegraphics{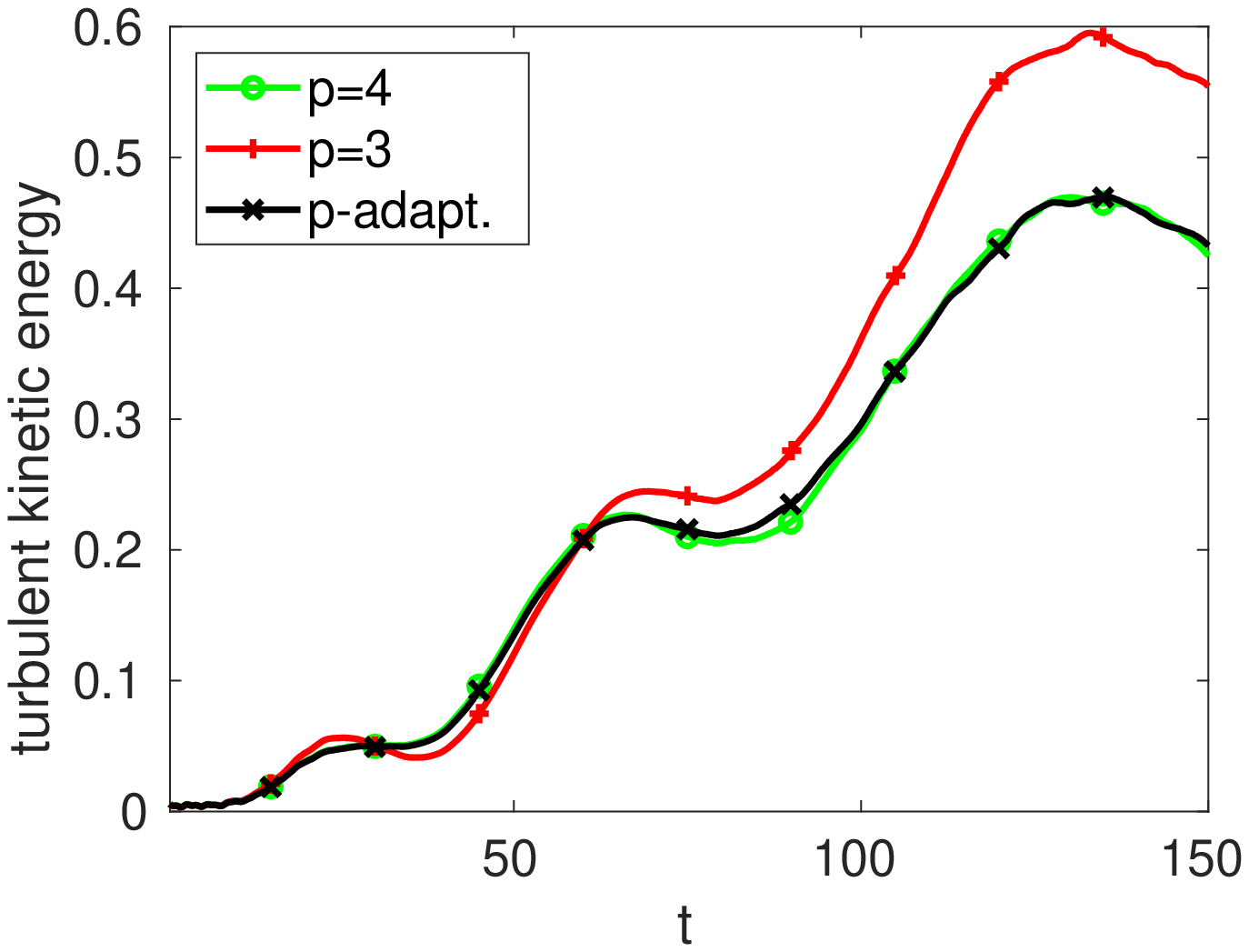}}
	    \label{fig:ml_kr}}
	\subfigure[]{\resizebox*{6cm}{!}
	    {\includegraphics{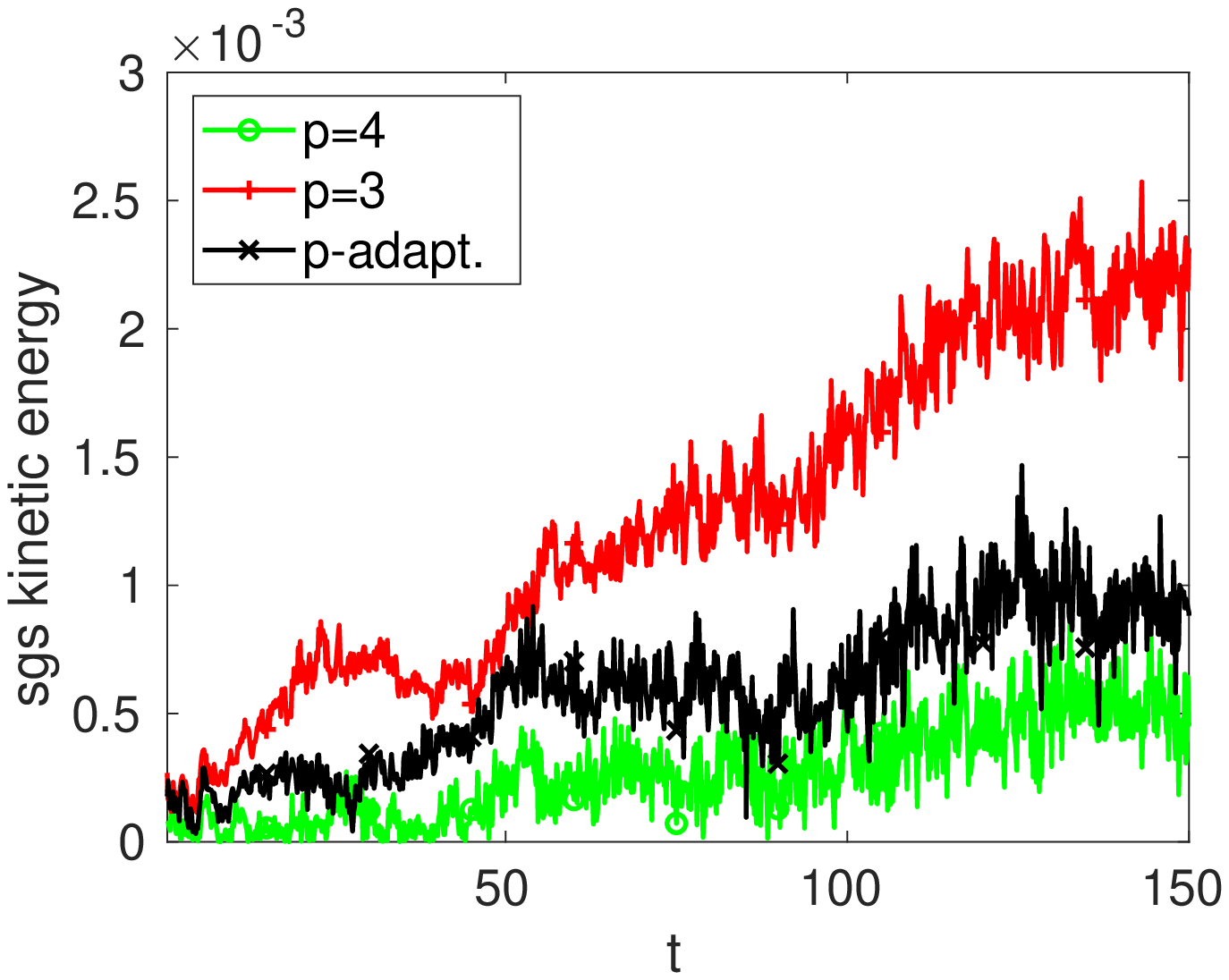}}
	    \label{fig:ml_tsgs}}
	\caption{Resolved turbulent kinetic energy (a) and SGS kinetic energy (b) in function of time for the mixing
	layer simulations.}
	\label{fig:ml_turb_en}
\end{figure}
Looking at the turbulent kinetic energy profile in Figure \ref{fig:ml_kr}, we can affirm that the turbulent energy amount of the flow is reproduced by the $p-$adaptive simulation with the same accuracy as in the  $p=4 $ case. 
In Figure \ref{fig:ml_tsgs}, the subgrid scale (SGS) kinetic energy in the  $p=3 $ case is clearly larger
 than in the other cases, due to the lower resolution that requires more intense contribution from the model. 
The SGS energy for the $p-$adaptive case is closer to that of the $p=4$ case, demonstrating that the two simulations have an equivalent effective resolution.

 \begin{figure}
	\centering
	\includegraphics[width=0.5\textwidth]{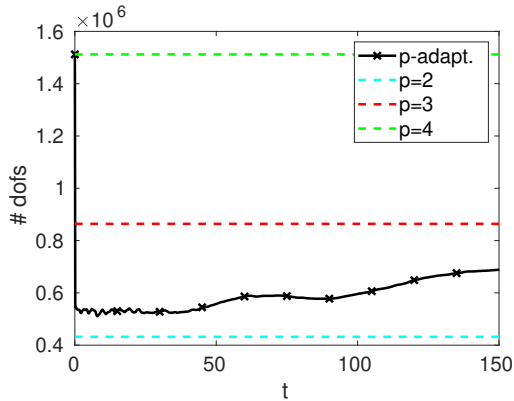}
	\caption{Total number of degrees of freedom as function of time for the mixing layer simulation: the behaviour of the 
	$p-$adaptive case is compared with values corresponding to the uniform degree  distributions $p=4$, $p=3$ and $p=2$. }
	\label{fig:ml_dof}
\end{figure}
All the simulations were carried out on the  Marconi cluster at CINECA, using 272 KNL processors.
In Figure \ref{fig:ml_dof}, the total number of dofs  for the adaptive simulation 
is shown as a function of time and compared with the values corresponding to uniform polynomial distributions,
 that are about $1.51\times 10^6$ and $8.64\times 10^5$
for $p=4$ and $p=3, $ respectively. The gain using $p-$adaptivity is evident, since it requires 
less than $80\%$ of the dofs for $p=3$ and less then half of the dofs for the $p=4$ case.
The core hours necessary to perform the simulations are reported in Table \ref{tab:ml_cputime}.
In spite of the absence of a specific dynamic load balancing procedure, the $p-$adaptive technique
also yields a  reduction  of  the required CPU time, which has a  value only marginally larger than that of the $p=3 $ simulation.
On the other hand, the sub-optimal nature of the present implementation is highlighted by the fact
that the  $p-$adaptive simulation required more CPU time than the $p=3 $ case, even though it involved less dofs.

\begin{table}[]
	\centering
	\begin{tabular}{|l|c|}
	\hline
		 & {Core hours} \\
	\hline
		$p= 4$     &  12600 \\
	\hline 
	        $p=3$    &   9000\\
	  \hline
	        $p-$adaptive  & 9300 \\
          \hline
	\end{tabular}
	\caption{Core hours required for the mixing layer simulations using uniform polynomial 
	degree distribution $p=4$, $p=3$ and $p-$adaptive, respectively.}
	\label{tab:ml_cputime}
\end{table}

Concluding this analysis, we can affirm that the  $p-$adaptive solutions display almost the same accuracy 
as those  obtained with constant degree four,  but require less then the half the number of dofs 
and about $29\% $ less CPU time for this test case, in spite of a sub-optimal implementation. 
Moreover, the structure function indicator which was proven successful in static adaptation in \cite{tugnoli:2017} is shown  to be suitable also for simulation of transitional flows using a dynamic adaptation in time.

\section{Interaction of Vortex and Square Section Cylinder}
\label{sec:Vortex_Cylinder}
A body-vortex interaction flow represents another interesting time dependent problem  to test dynamic $p-$adaptivity.
Here, a vortex interacting with a square section cylinder at $Re=22000$ and $Ma=0.3$ is considered. 
This configuration was chosen because the analogous case without a vortex has already been studied
with statically adaptive techniques in  \cite{tugnoli:2017}.
The classical Smagorinsky SGS model was applied for this test.
The results obtained with polynomial adaptivity are compared with the solution obtained with  uniform 
in space and constant in time degree 
distribution. The  constant degree four solution is taken as reference. 

\begin{figure}
  \centering
  \includegraphics[width=.6\textwidth]{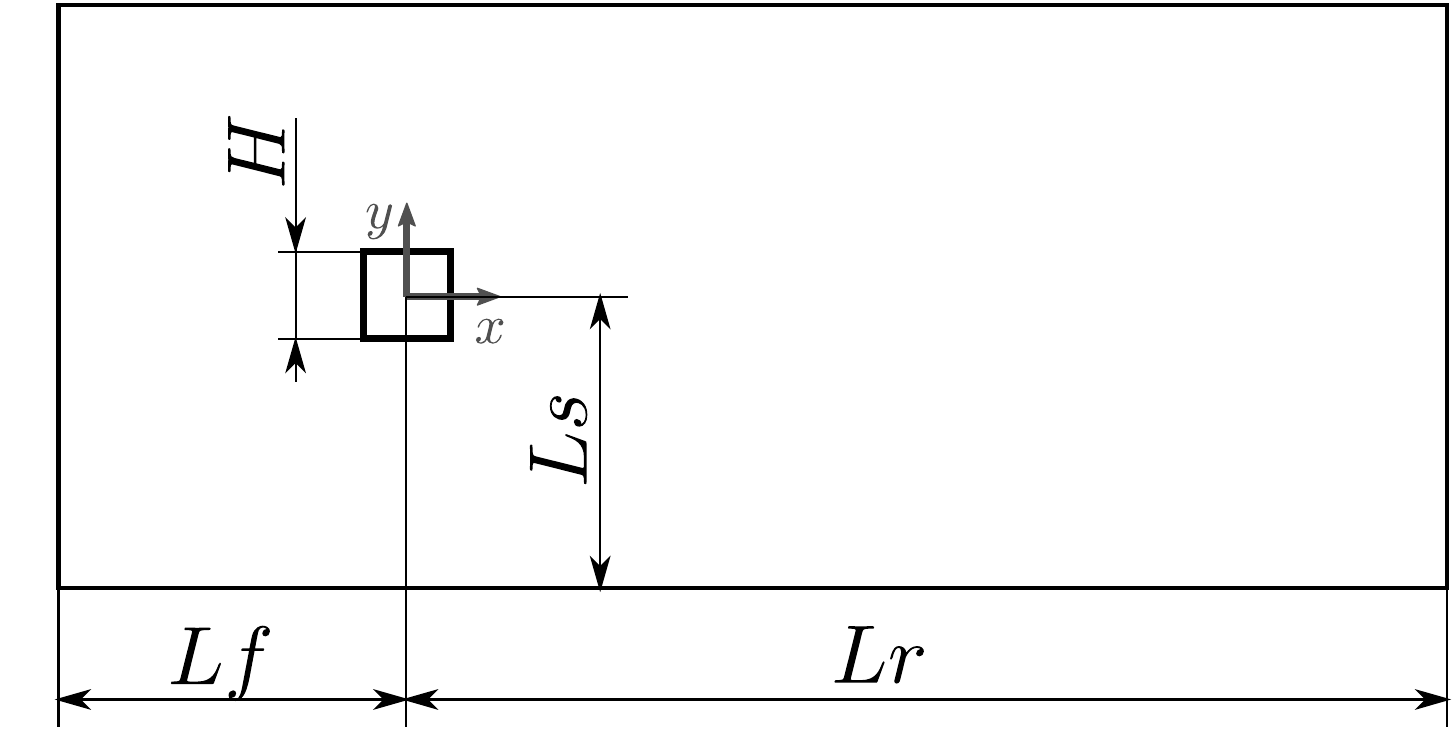}
  \caption{2D sketch of the square section cylinder case geometry. The lines on which the statistics are calculated and 
  plotted are highlighted.}
  \label{fig:cyl-geometry}
\end{figure}

The computational domain employed is box shaped and a 2D sketch of the geometry is represented in 
Figure~\ref{fig:cyl-geometry}. 
Denoting the cylinder side with $H$, which is used as reference length, the inflow length has been taken equal to 
$L_f=10H,$ while the outflow length is equal to $L_r=20H$. The cylinder is vertically centred with a distance of 
$L_s=10H$ from the upper and lower boundaries.
 The domain is extruded in the spanwise direction by a length equal to $L_z=4H$.
  This configuration was employed e.g. in  \cite{colonius:1991} and has been used 
 to assess statically adaptive  LES simulations of the standard flow around a square section cylinder
 in \cite{tugnoli:2017, tugnoli:2017a, tugnoli:2019}.
 
Non-slip isothermal conditions are imposed on the cylinder walls, while Neumann conditions  are imposed on the upper and lower boundaries of  the domain and periodic conditions are enforced in the spanwise direction to simulate an infinite span cylinder. 
Dirichlet conditions with the far field values are imposed on the inflow and outflow, with sponge layers to avoid 
reflections of disturbances from the boundaries. At the inflow, a uniform far-field velocity is imposed. 

The mesh employed consists of 23816 tetrahedra arranged in an outer unstructured area and a structured O-grid mesh around 
the cylinder, which is then  extruded in the spanwise direction. 
Considering a polynomial basis of degree 4, the resolution in space around the cylinder is $\Delta_n = 0.008H$ in the wall 
normal direction, $\Delta_t = 0.024H$ along the cylinder sides and $\Delta_z = 0.135$ along the spanwise direction.
As in \cite{tugnoli:2017}, the thresholds for the structure function (SF) indicator are  taken to be $\epsilon_1 = 5\times 10^{-4}, $   $\epsilon_2  =  1\times 10^{-2}, $ so as  to obtain  a   number of dofs just lower than the one obtained with uniform degree 3. 

\begin{figure}
  \centering
  \includegraphics[width=.6\textwidth]{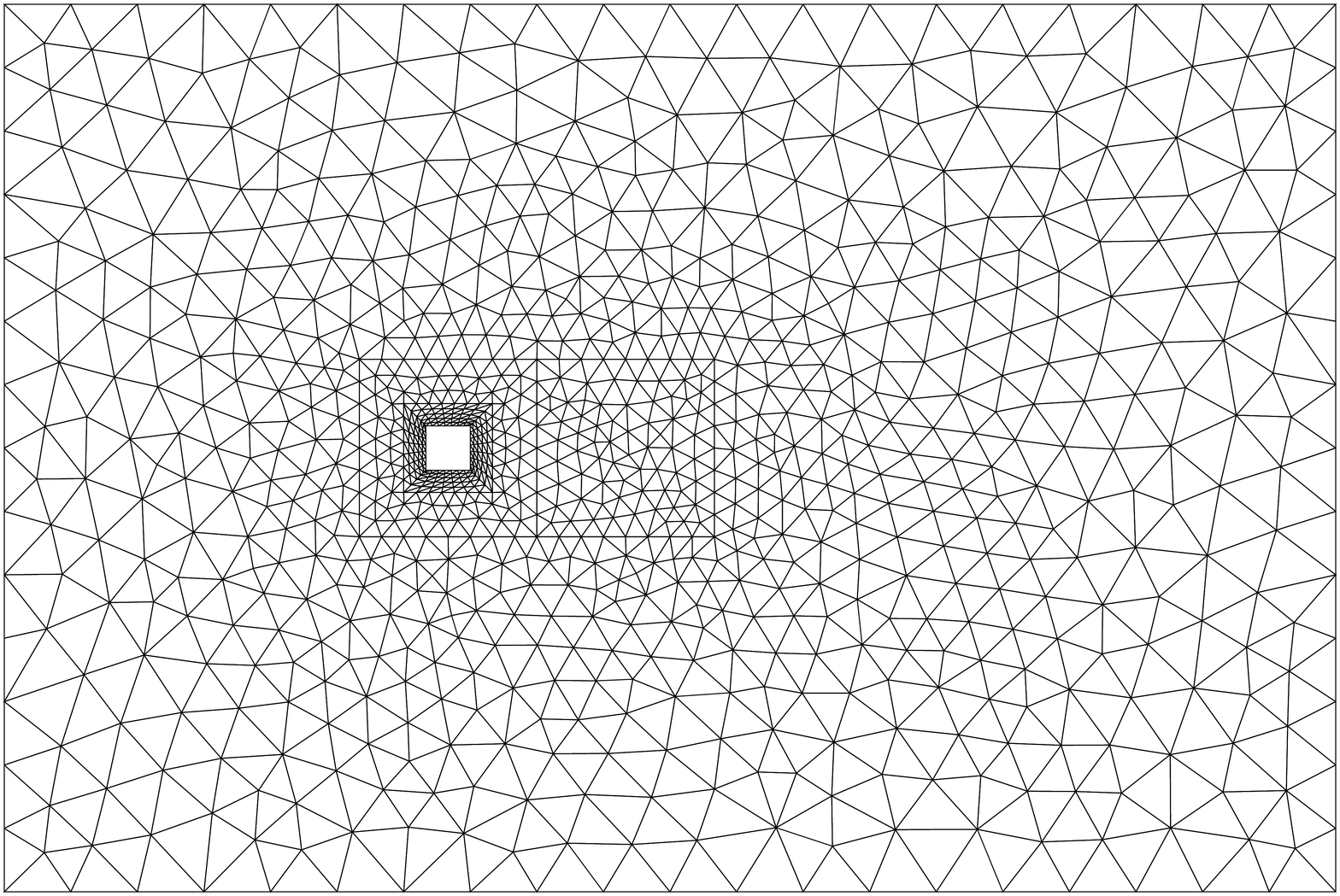}
  \caption{2D section of the mesh employed for the flow around the square section cylinder}
  \label{fig:cyl-mesh}
\end{figure}

To simulate the cylinder-vortex interaction, a vortex is superimposed
to an initial condition corresponding to a developed 
flow around the cylinder, obtained by a previous LES simulation.  
The introduced vortex is parallel to the $z$ direction and defined by the velocity components
\begin{eqnarray}
u &=& U_\infty - C_v \exp{\left(-\frac{x^2+y^2}{2R_v^2}\right)},\\
v &=& C_v \exp{\left(-\frac{x^2+y^2}{2R_v^2}\right)},\\
w &=& 0,
\end{eqnarray}
where $C_v$  denotes the vortex strength, $R_v$ is the vortex radius, $r^2 = x^2+y^2$ 
is the distance in the $x-y$ plane from the vortex 
centre and $U_\infty$ is the uniform flow over which the vortex is superimposed \cite{lodato:2008}. 
The maximum radial velocity is obtained at $r = R_v$ and is given by $v_{\theta_{max}} = \frac{C_v}{R\sqrt{e}}$.
Regarding the other variables,  it is known from \cite{colonius:1991}  that for a viscous compressible vortex the radial 
pressure distribution given by the solution of
\begin{equation}
\partfrac{p}{r} = \frac{\rho u_\theta^2}{r}.
\label{eq:radial_p_ode}
\end{equation}
Assuming a constant temperature distribution $T = T_\infty, $ one obtains from \eqref{eq:radial_p_ode} 
the profile
\begin{equation}
P = P_\infty \exp{\left[ - \frac{C_v^2}{2 R_{gas} T_\infty R_v^2} \exp{\left( - \frac{r^2}{R_v^2}\right)} \right]}.
\end{equation}
Finally the density distribution follows the equation of state
$ \rho = P/(R_{gas} T). $

In \cite{tugnoli:2017a, tugnoli:icosahom}  simple advection of this vortex was tested. 
The maximum radial velocity chosen is $v_{\theta_{max}} = 0.5$ and the vortex radius is $R_v = 0.41$.  
 A reference simulation without vortex was used to calibrate the introduction time in order for the vortex to reach the cylinder in the instant of maximum lift.
The dynamical $p-$adaptation strategy consisted in this case of adapting the polynomial degrees approximately three times
during the time needed by a fluid particle to pass through the smallest elements on the cylinder walls at the free stream velocity $U_\infty$. This choice led to  computing the indicator value at intervals
$\Delta t_i=0.0016 $ and averaging them over intervals  
 $\Delta t_a= 0.05, $ while the value  $\Delta t=0.0004 $ was used for time integration.

\begin{figure}
	\centering
\includegraphics[width=0.8\textwidth]{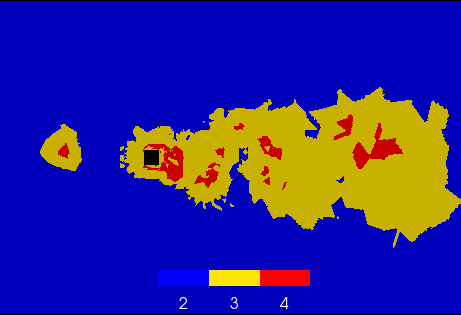} (a)
		  \includegraphics[width=0.8\textwidth]{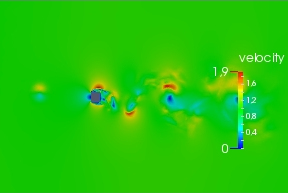}(b)
	\caption{Polynomial degree (a) and momentum magnitude fields (b), for the 
	square cylinder--vortex interaction flow at time $t=0.25.$}
	\label{fig:vortcyl_degs_momentum_t025}
\end{figure}

\begin{figure}
	\centering
\includegraphics[width=0.8\textwidth]{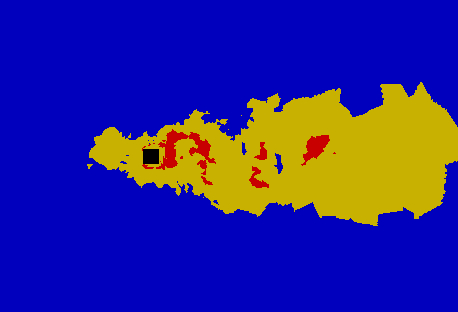} (a)
		  \includegraphics[width=0.8\textwidth]{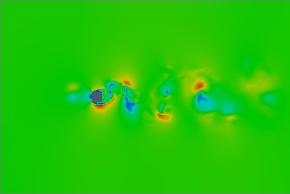}(b)
	\caption{Polynomial degree (a) and momentum magnitude fields (b), for the 
	square cylinder--vortex interaction flow at time $t=4.$}
	\label{fig:vortcyl_degs_momentum_t4}
\end{figure}

\begin{figure}
	\centering
\includegraphics[width=0.8\textwidth]{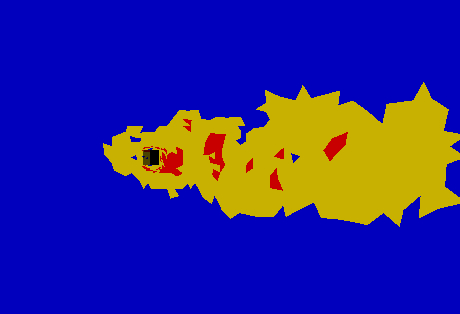} (a)
		  \includegraphics[width=0.8\textwidth]{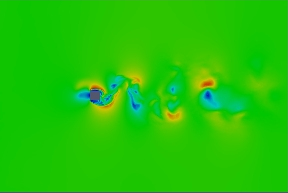}(b)
	\caption{Polynomial degree (a) and momentum magnitude fields (b), for the 
	square cylinder--vortex interaction flow at time $t=5.$}
	\label{fig:vortcyl_degs_momentum_t5}
\end{figure}

\begin{figure}
	\centering
\includegraphics[width=0.8\textwidth]{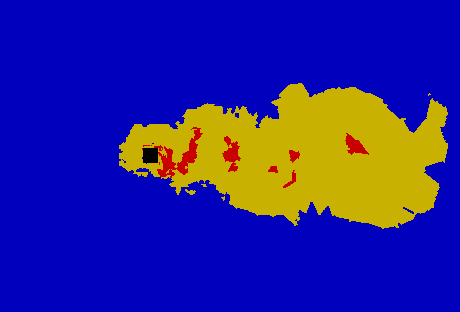} (a)
		  \includegraphics[width=0.8\textwidth]{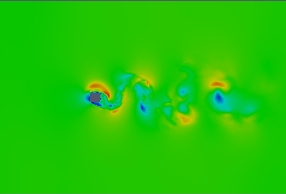}(b)
	\caption{Polynomial degree (a) and momentum magnitude fields (b), for the 
	square cylinder--vortex interaction flow at time $t=6.$}
	\label{fig:vortcyl_degs_momentum_t6}
\end{figure}

 The dynamically adaptive procedure was able also in this case  to effectively represent the structures of the flow, both in the 
advected vortex region and in the wake of the obstacle, as can be seen in Figures \ref{fig:vortcyl_degs_momentum_t025}-\ref{fig:vortcyl_degs_momentum_t6}.
It is possible to note that a higher polynomial degree is employed in the advected vortex region and 
in the shear layers around the cylinder. The dynamic adaptation procedure is able to effectively increase the polynomial degree 
around the vortex and to follow it as it is advected downstream, leaving all the 
elements with no vortex activity at the lowest  resolution. 
Notice that the centre of the vortex is represented with  degree 4 polynomials in the first instants, but only degree 3 polynomials
are employed later, since the vortex enters an area with a more refined mesh in which a lower polynomial degree
is sufficient.

\begin{figure}
	\centering
	\subfigure[Lift coefficient]{\resizebox*{6cm}{!}
		{\includegraphics{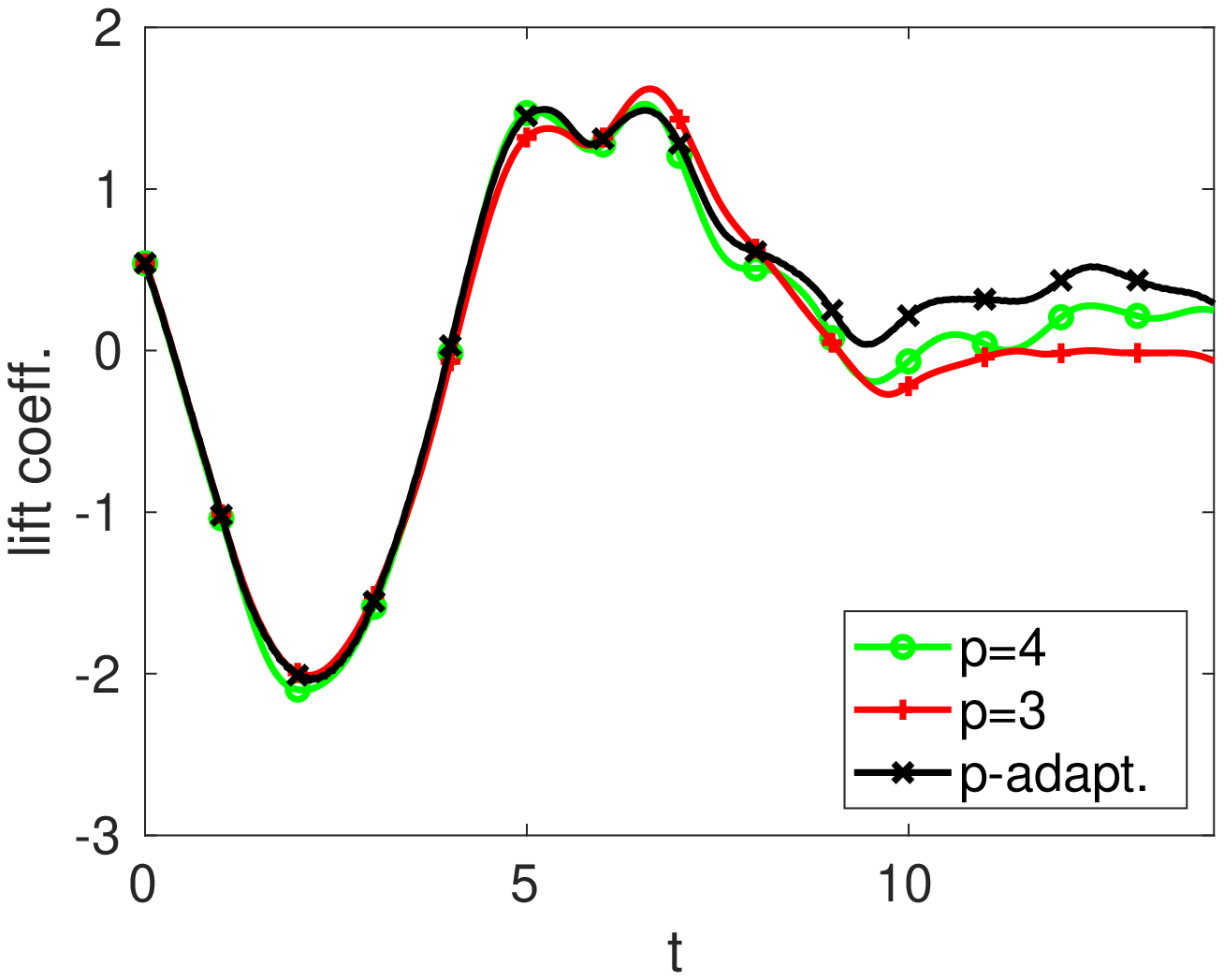}}
		\label{subfig:lmax_cl}}
	\hfill
	\subfigure[Drag coefficient]{\resizebox*{6cm}{!}
		{\includegraphics{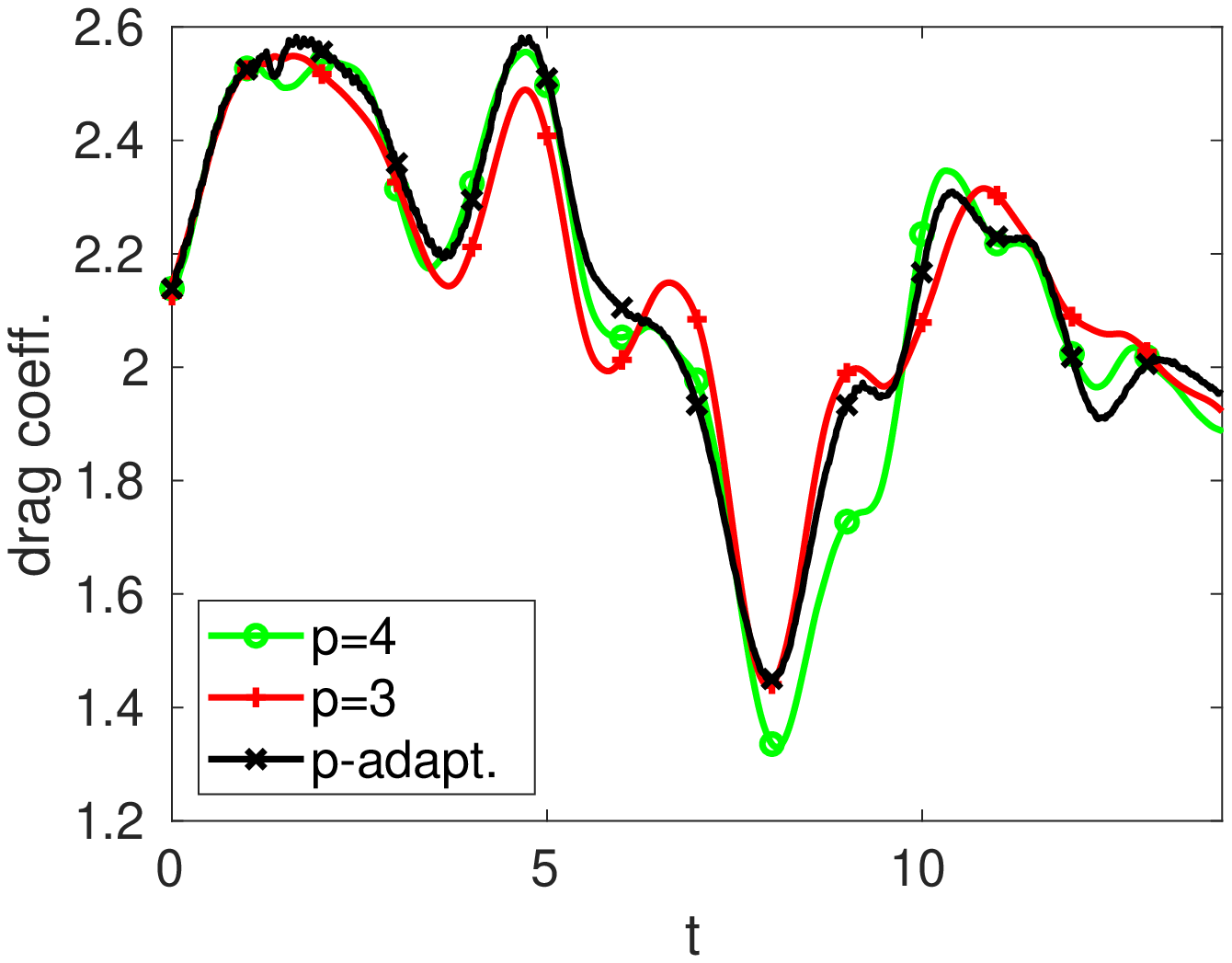}}
		\label{subfig:lmax_cd}}
	\caption{Forces coefficients in function of time for vortex interacting with the cylinder.}
	\label{fig:vortcyl_Lmax_forces}
\end{figure}

Among the different effects caused by the vortex-cylinder interaction,
 the most interesting are those on the forces acting on the 
cylinder. The history of the load coefficients before, during and after the interaction is
reported in Figure \ref{fig:vortcyl_Lmax_forces}.

In the developed flow around the square cylinder, the periodic oscillation of the force coefficients are related 
to the unsteadiness of the recirculation bubble on the upper and lower sides and to the vortex shedding. Reaching 
the upper corner of the cylinder at $t=5$, the vortex changes the pressure distribution on the front of the cylinder and 
strongly modifies the recirculating bubble on the upper side. In this condition, it is very 
important to use an adequate resolution on the upper side and in the separation bubble. 
In Figure \ref{fig:vortcyl_Lmax_forces} it can be observed that the peaks of the lift and drag coefficients at $t=5$ with 
$p-$adaptivity equal those in the reference solution $p=4,$ demonstrating the capability of the proposed dynamically 
adaptive approach to well represent this unsteady phenomenon.
The values of the differences of the 
peaks for the lift and drag coefficients with respect to the vortex free solution are reported in Table \ref{tab:vortcyl-peaks}.
Also in this case the values obtained with $p-$adaptivity are in very good agreement with the reference solution. 

\pagebreak

\begin{table}[]
	\centering
	\begin{tabular}{|l|c|c|c|}
	\hline
		configuration & {$C_L$ peak diff. \%} & {$C_D$ max diff. \%} & {$C_D$ min diff. \%} \\
	\hline
		$p= 4$     & -24  & -0.01 & -39 \\
	\hline 
	        $p=3$    &   -32  &  -4 & -38\\
	  \hline
	        $p-adaptive$  & -24    &  -0.01  & -38 \\
          \hline
	\end{tabular}
	\caption{Differences, in percentage, of the force coefficients with respect to the simulation without vortex, 
	during the vortex interaction time, $t=4-8.5$. For the lift, only the change in the subsequent peak is presented, 
	for the drag both the difference in maximum and minimum.}
	\label{tab:vortcyl-peaks}
\end{table}

\begin{figure}
	\centering
	\includegraphics[width=.5\textwidth]{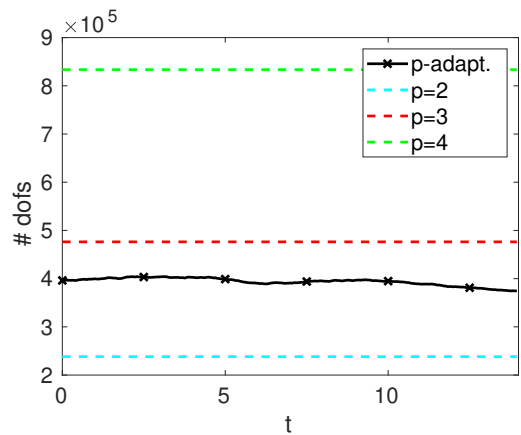}
	\caption{Number of degrees of freedom during the dynamic adaptive simulation of vortex-cylinder interaction: $p-$ adaptive
	simulation compared with uniform polynomial distributions $p=2$, $p=3$ and $p=4$.}
	\label{fig:vortcyl_nelems}
\end{figure}

The simulations were carried out on the  Marconi cluster at CINECA, using 272 Brodwell processors.
The adaptive simulation allows a  reduction of about $60\%$  of the required CPU time with respect to the constant $p=4$ degree case.
 In Figure \ref{fig:vortcyl_nelems}, the time history of the dofs number is shown. It is worth noting that,
  even though the number of dofs is affected by the   flow regime inside the domain, 
its variations are anyway fairly limited and the total number remains less than the one employed for the uniform $p=3$
simulation. 
The core hours necessary to perform the simulations are reported in Table \ref{tab:vort_cputime}.
Also in this case,the $p-$adaptive technique
 yields a  reduction  of  the required CPU time, which has a  value only marginally larger than that of the $p=3 $ simulation.
Again, the sub-optimal nature of the present implementation is highlighted by the fact
that the  $p-$adaptive simulation required more CPU time than the $p=3 $ case, even though it involved less dofs.

\begin{table}[]
	\centering
	\begin{tabular}{|l|c|}
	\hline
		 & {Core hours} \\
	\hline
		$p= 4$     & 756 \\
	\hline 
	        $p=3$    &  388\\
	  \hline
	        $p-$adaptive  &  462 \\
          \hline
	\end{tabular}
	\caption{Core hours required for the vortex-cylinder simulations using uniform polynomial 
	degree distribution $p=4$, $p=3$ and $p-$adaptive, respectively.}
	\label{tab:vort_cputime}
\end{table}

\section{Conclusion}
\label{sec:Conclu}

The need  for adaptive  LES approaches was first stated in \cite{pope:2004},
but adaptive LES simulations are not very common yet. DG methods provide a appropriate environment 
for adaptive approaches, since they do not require to enforce continuity constraints at the inter-element boundaries.
Furthermore, $p-$adaptive techniques are appealing, since they allow to correct possible  shortcomings of the computational
mesh as well as to perform dynamically adaptive simulations without extensive remeshing.

Static polynomial adaptivity has been applied to LES in a DG context in 
\cite{delallaveplata:2019, tugnoli:2017, naddei:2019,  tugnoli:2017a}. 
An essential tool for such simulations
is an adaptation criterion that, rather than simply increasing the resolution in order to decrease the error,
which is known to  lead to a DNS solution \cite{mitran:2001, sagaut:2006},  tries instead to adjusting the resolution in order to directly resolve only a prescribed amount of the  turbulent scales.

In this work, the physically based refinement indicator proposed by the authors in \cite{tugnoli:2017, tugnoli:2017a} was shown to be applicable locally and  dynamically in time, in order to simulate efficiently and accurately transient phenomena. 
Numerical simulations of a temporally evolving mixing layer and of   a vortex impinging on a square cylinder   were performed.
The results obtained show that a significant reduction of the computational cost of LES is achieved, reducing the number of degrees of freedom employed to values down to $50\% $ of those required by constant maximum polynomial degree simulations, while maintaining essentially the same level of accuracy.
   
A main limitation of the present results is that, in this preliminary implementation of a dynamically adaptive approach, no dynamic load balancing has been performed.  
In future work on code optimization,  load balancing approaches such as those employed in \cite{bassi:2019,wang:2019} will be considered and tested in order to achieve maximum parallel efficiency.

\section*{Acknowledgements}
We acknowledge that the results of this research were made possible by the  computational resources made available   
at CINECA (Italy) by the high performance computing projects ISCRA-C HP10CHV1QD and ISCRA-C HP10C6F9BK.

\bibliographystyle{plain}
\bibliography{dg_les}

\end{document}